\documentclass[letterpaper,10pt]{article}
\usepackage{amsmath,amssymb}
\usepackage{graphicx}
\usepackage[super]{cite}
\usepackage[hmargin=1in,vmargin=1in]{geometry}

\newcommand{\mb}[1]{\mathbf{#1}}
\newcommand{\bs}[1]{\boldsymbol{#1}}

\allowdisplaybreaks[4]

\begin{document}

%\allsectionsfont{\normalsize}

\title{End-monomer Dynamics in Semiflexible Polymers}

\author{Michael Hinczewski\thanks{Feza G\"ursey Research Institute, T\"UBITAK -
Bosphorus University, \c{C}engelk\"oy 34684, Istanbul, Turkey} , Xaver Schlagberger\thanks{Physics Department, Technical University of Munich,  85748 Garching, Germany} ,  Michael Rubinstein\thanks{Department of Chemistry, University of North Carolina, Chapel Hill, NC 27599-3290, USA} ,\\ Oleg Krichevsky\thanks{Physics Department, Ben-Gurion University, Beer-Sheva 84105, Israel} , and Roland R. Netz$^\dagger$}

\date{}

\maketitle
\begin{abstract}

Spurred by an experimental controversy in the literature, we
investigate the end-monomer dynamics of semiflexible polymers through
Brownian hydrodynamic simulations and dynamic mean-field theory.
Precise experimental observations over the last few years of
end-monomer dynamics in the diffusion of double-stranded DNA have
given conflicting results: one study indicated an unexpected
Rouse-like scaling of the mean squared displacement (MSD) $\langle
r^2(t) \rangle \sim t^{1/2}$ at intermediate times, corresponding to
fluctuations at length scales larger than the persistence length but
smaller than the coil size; another study claimed the more
conventional Zimm scaling $\langle r^2(t)\rangle \sim t^{2/3}$ in the
same time range.  Using hydrodynamic simulations, analytical and
scaling theories, we find a novel intermediate dynamical regime where
the effective local exponent of the end-monomer MSD, $\alpha(t) =
d\log\langle r^2(t) \rangle /d\log t$, drops below the Zimm value of
2/3 for sufficiently long chains.  The deviation from the Zimm
prediction increases with chain length, though it does not reach the
Rouse limit of 1/2.  The qualitative features of this intermediate
regime, found in simulations and in an improved mean-field theory for
semiflexible polymers, in particular the variation of $\alpha(t)$ with
chain and persistence lengths, can be reproduced through a heuristic
scaling argument.  Anomalously low values of the effective exponent
$\alpha$ are explained by hydrodynamic effects related to the slow
crossover from dynamics on length scales smaller than the persistence
length to dynamics on larger length scales.

\end{abstract}

\section{Introduction}\label{intro}

Recent experimental advances using fluorescence correlation
spectroscopy \cite{Lumma,Shusterman,Winkler,Bernheim,Petrov} have
given unprecedented information about the dynamical behavior of large
single polymer molecules in solution, in particular the small-scale
kinetics of individual monomers inaccessible to traditional techniques
like dynamic light scattering. One of the first studies along this
direction yielded an unexpected result. Shusterman {\it et al.}
\cite{Shusterman} observed the random motion of a single labeled
monomer at the end of a long double-stranded DNA molecule, and found
evidence of an ``intermediate Rouse regime'': the mean squared
displacement (MSD) followed a scaling $\langle r^2(t)\rangle \propto
t^{1/2}$ for a wide time range corresponding to polymer motion at
length scales smaller than the coil size $R_g$ but larger than the
persistence length $l_p$.  This agrees with the free-draining Rouse
model for a polymer which neglects hydrodynamic interactions mediated
by flow fields arising from the monomers moving through the solvent.
Such a result contradicts the conventional wisdom for flexible
polymers, which states that these hydrodynamic interactions play a
crucial role in polymer dynamics in dilute solutions and give rise to
non-draining behavior that is qualitatively described by the Zimm
theory, which predicts $\langle r^2(t)\rangle \propto
t^{2/3}$~\cite{Zimm,DoiEdwards}.  Though double-stranded DNA is a
semiflexible polymer (having a persistence length $l_p \approx 50-100$
nm much larger than the width $\approx 2$ nm), the expectation for
kinetics at scales larger than $l_p$ is that it behaves like a
non-draining flexible polymer. Thus the apparent absence of
hydrodynamic effects is quite surprising, and the intermediate Rouse
regime does not fit into established theories of the dynamics of
\emph{flexible} polymers in dilute solutions, though recently there
has been an attempt to explain its existence through a theory
exhibiting time-dependent hydrodynamic screening \cite{Lisy}.  On the
other hand, a new experimental study by Petrov {\it
  et. al.}~\cite{Petrov} on the same system did not seem to show the
Rouse regime, and its results were interpreted to be generally
consistent with the dynamics predicted by the Zimm theory. Arguably
the dynamics of a semiflexible polymer such as DNA may be expected to
differ from that of flexible polymers. However, with the exception of
the Harnau, Winkler, Reineker (HWR) model~\cite{HWR2}, other
established theories of the dynamics of semi-flexible
polymers~\cite{Kroy,Granek} treat only the range of displacements
smaller than the persistence length $\langle r^2(t)\rangle < l_p^2$.

To help resolve the controversy over the dynamics of semiflexible
polymers on intermediate length scales $l_p^2 < \langle r^2(t) \rangle
< L^2$, we study the end-monomer behavior of semiflexible chains in
dilute solutions using two approaches: dynamic mean-field theory (MFT)
that includes hydrodynamics with the pre-averaging approximation, and
Brownian hydrodynamics simulations without the pre-averaging
approximation.  The end-monomer MSD, diffusion constants, and longest
relaxation times from the two approaches agree closely with each
other.  While the hydrodynamic pre-averaging MFT method is similar to
that of HWR in Ref.~\citen{HWR2}, we have improved the approximation
by taking into account the full hydrodynamic interaction matrix in the
Langevin equation, and not just the diagonal contribution.  This leads
to much better agreement between the MFT and the simulation data:
compared with the earlier version, the improved MFT is 10--65\% closer
to the mean-square displacement $\langle r^2(t)\rangle$ of the end
monomer obtained by simulations for time scales shorter than the
longest relaxation time of the chain, and reduces the discrepancy in
the effective local exponent $\alpha(t) = d\log\langle r^2(t)
\rangle/d\log t$ in this time range, which is underestimated by as
much as 10\% using the earlier method.  Thus we can confidently extend
the MFT to larger chain lengths $L$ that are inaccessible to
simulation.  For these chains we find an intermediate dynamical regime
where the continuously varying effective local exponent of the
end-monomer MSD, $\alpha(t)$, drops below 2/3, and its difference from
this Zimm value increases with $L$.  The existence of this regime and
the qualitative trends of $\alpha(t)$ with changing $L$ and $l_p$ are
verified independently through a heuristic scaling argument.  However
even at the largest chain lengths examined, comparable to or longer
than the experimentally studied chains of Refs.~\citen{Shusterman} and
\citen{Petrov}, the effective exponent $\alpha(t)$ does not reach the
Rouse limit of 1/2.  Comparison with the experimental MSD data of
Ref.~\citen{Shusterman} reveals two interesting results: the MFT
accurately describes the long-time diffusion behavior, related to the
large-scale dynamics of the chain; however at shorter times it
underestimates the extent of the MSD.  As we show in this paper, the
same sub-Zimm scaling of the MSD is also contained in the HWR theory
that was used to successfully fit the data of Ref.~\citen{Petrov}.  So
the question is not whether an intermediate sub-Zimm scaling regime
exists, but rather how large that regime is and how small are the
intermediate exponents.  The remaining discrepancy between theory and
experiment discussed in this paper highlights the importance of
additional dynamical degrees of freedom absent in the worm-like chain
model used as the starting point for the theoretical description of
DNA or some shortcomings in the current analysis of FCS data.

The paper is organized as follows: in Sec.~\ref{scaling} we give a
heuristic scaling argument that captures the basic properties of the
intermediate dynamical regime; in Sec.~\ref{brownian} we describe the
details of the Brownian dynamics simulations; in Sec.~\ref{mft} we
give an overview of the mean-field model for semiflexible polymers and
the pre-averaging approximation used to determine its behavior in
solution; in Sec.~\ref{results} we compare the simulation, MFT, and
heuristic results, together with the experimental data.  The dynamical
regimes exhibited in these results are examined through asymptotic
scaling analysis in Sec.~\ref{discussion}, and placed in the context
of earlier theories.  Finally Sec.~\ref{conclusion} summarizes the
main points of the paper.  Additional material, extending the
mean-field model of Sec.~\ref{mft} to extensible worm-like chains, is
provided in Appendix A.  Mathematical details of an analytical
approximation used in Sec.~\ref{discussion} are given in Appendix B.

\section{Heuristic Scaling Argument}\label{scaling}

\begin{figure}
\begin{center}
\includegraphics*{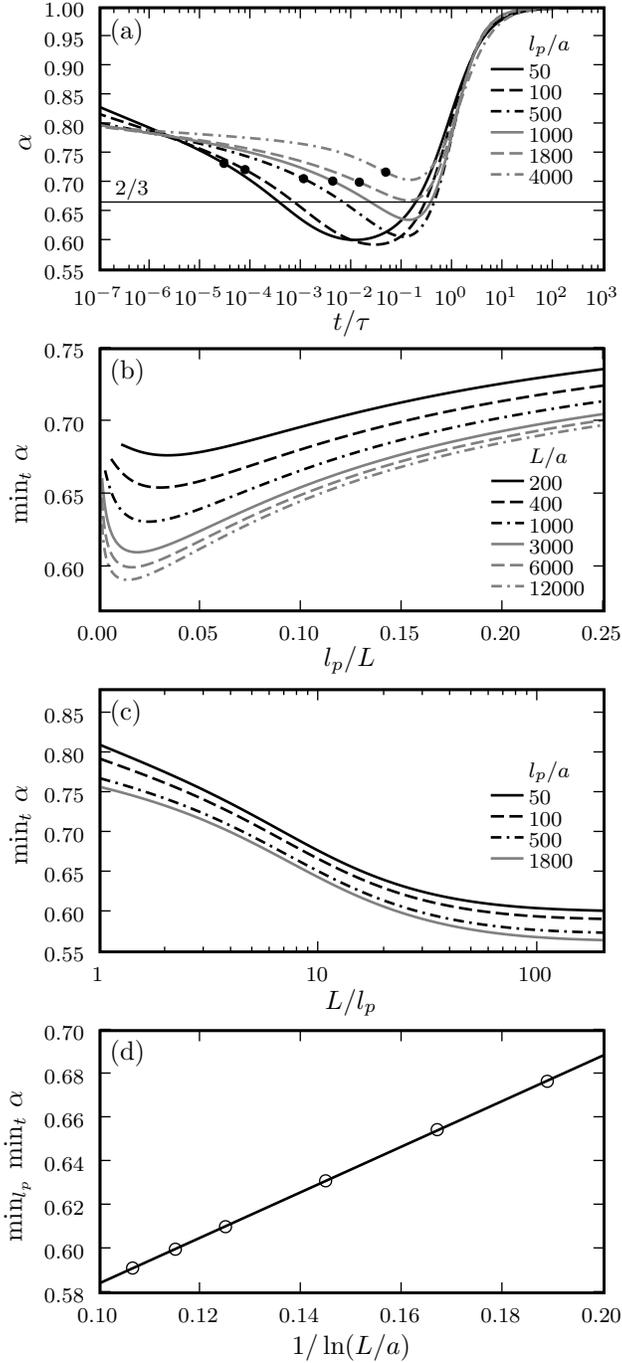}
\end{center}
\caption{Results of the heuristic scaling argument described in
  Sec.~\ref{scaling}.  (a) The effective exponent $\alpha = d \log
  \langle r^2(t) \rangle /d \log t$ versus $t/\tau$ for polymers of
  diameter $2a$, contour length $L = 12000a$, and various persistence
  lengths $l_p = 50a - 4000a$.  The dot along each $\alpha(t)$ curve
  marks $\alpha(\tau_p)$.  The time scales $\tau$ and $\tau_p$ are
  defined in the text.  (b) The minimum value of $\alpha$ over all $t$
  versus $l_p/L$ for several $L = 200a - 12000a$.  (c) The minimum
  value of $\alpha$ over all $t$ versus $L/l_p$ for several $l_p = 50a
  - 1800a$.  (d) The circles show the minimum value of $\alpha$ in
  each of the curves in panel (b), plotted as a function of
  $1/\ln(L/a)$.  Superimposed is a straight-line fit to the data
  points, $\min_{l_p} \min_t \alpha \approx 0.48+1.04/\ln(L/a)$.}
\label{fig:heur}
\end{figure}

\begin{figure}
\begin{center}
\includegraphics*{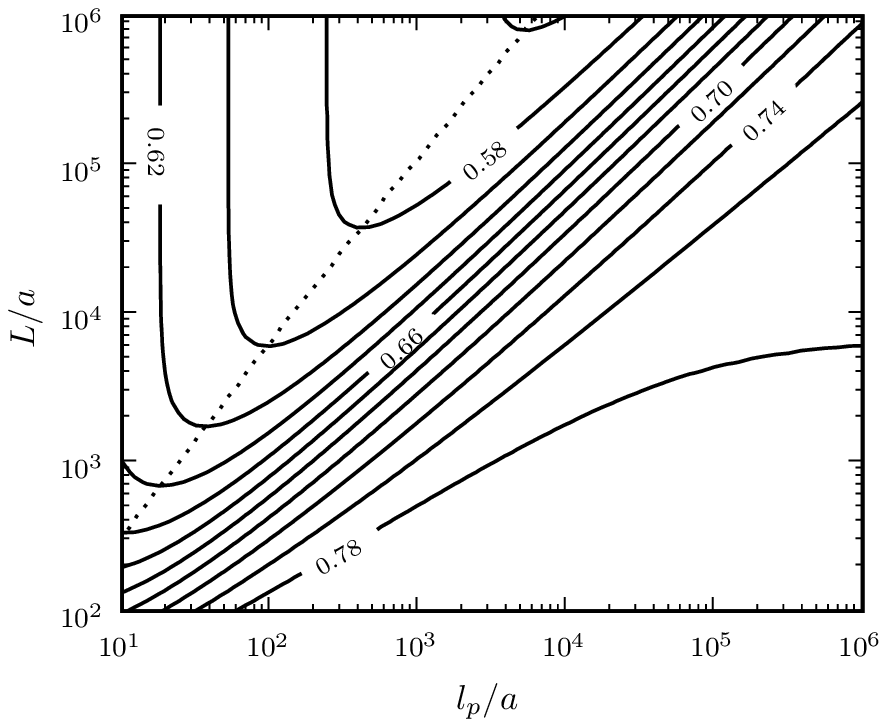}
\end{center}
\caption{The contours show the minimum value of the effective exponent
  $\alpha(t) = d \log \langle r^2(t) \rangle /d \log t$ over all $t$
  calculated using the heuristic scaling argument in
  Sec.~\ref{scaling} for chains of diameter $2a$, total length $L$,
  and persistence length $l_p$.  The interval between contours is
  0.02.  The dotted line indicates the minimum value of $\min_t
  \alpha$ over $l_p/a$ for a given $L/a$.}
\label{fig:contours}
\end{figure}

Certain qualitative features of the intermediate dynamical regime for
semiflexible polymers can be derived from a scaling argument similar
to the one that is typically used to understand subdiffusive motion in
the Rouse or Zimm models \cite{RubinsteinColby}. In our heuristic
scaling, we assume that at time scales $t$ the diffusion of the
polymer is characterized by the coherent motion of a section with
contour length $\ell(t)$. The MSD of a monomer during this time $t$
has two limiting subdiffusive behaviors depending on the magnitude of
$\ell(t)$.  For $\ell(t) \ll l_p$, where the local stiffness of the
polymer plays the key role, the MSD $\langle r^2(t) \rangle$ is
dominated by transverse chain fluctuations and thus $\langle r^2(t)
\rangle \sim \ell(t)^3/l_p$
\cite{Kroy,Granek,Morse,GittesMackintosh,Everaers,RubinsteinColby}.
For $\ell(t) \gg l_p$, the MSD can be estimated as $\langle r^2(t)
\rangle \sim r^2_\text{ee}(\ell(t))$, where $r^2_\text{ee}(\ell)$ is
the mean squared end-to-end distance of a polymer of contour length
$\ell$.  For a semiflexible chain with persistence length $l_p$ this
is given by:
\begin{equation}\label{eq:s1}
r^2_\text{ee}(\ell) = 2 l_p\ell - 2l_p^2(1-e^{-\ell/l_p})\,. 
\end{equation}
Thus the total intramolecular contribution to the MSD, $\langle
r^2_\text{intra}(t) \rangle$, can be written in a heuristic form which
smoothly interpolates between these two limits:
\begin{equation}\label{eq:s1b}
\langle r^2_\text{intra}(t) \rangle = C_1 \left[
  \{r^2_\text{ee}(\ell(t))\}^{-\phi_1} + C_1^\prime\left\{\frac{(r^2_\text{ee}(\ell(t)))^{3/2}}{l_p}\right\}^{-\phi_1} 
  \right]^{-1/\phi_1}\,.
\end{equation}
Using the fact that $r^2_\text{ee}(\ell) \approx \ell^2$ for $\ell \ll
l_p$, one can easily check that Eq.~\eqref{eq:s1b} has the appropriate
asymptotic limits for small and large $\ell(t)$, with the crossover
occurring at $t \approx \tau_p$ where the relaxation time $\tau_p$ of
a persistent segment is defined by $r^2_\text{ee}(\ell(\tau_p)) =
l_p^2$.  We added numerical constants $C_1,\: C_1^\prime \sim
\text{O}(1)$ and a crossover exponent $\phi_1$ which will be
determined from the comparison with numerical data.

In order to close the scaling argument, we write the diffusion
relation between spatial and temporal scales 
\begin{equation} \label{diffusion}
\langle r^2_\text{intra}(t) \rangle = 6 D(\ell(t)) t,
\end{equation}
in terms of an effective diffusion constant $D(\ell)$ for a polymer
section of length $\ell$.  $D(\ell)$ can be estimated from a heuristic
formula~\cite{Schlagberger} that compares well with both simulation
and experimental DNA results:
\begin{equation}\label{eq:s2}
D(\ell) = C_2(D_\text{cyl}^{\phi_2}(\ell) + D_\text{coil}^{\phi_2}(\ell))^{1/\phi_2}\,,
\end{equation}
where
\begin{equation}\label{eq:s3}
\begin{split}
D_\text{cyl}(\ell) &= \frac{2\mu_0 k_B T}{\ell}\left[\log \frac{\ell}{2a} + 0.312\right.\\
&\qquad\qquad\quad\left. +0.565 \left(\frac{2a}{\ell}\right)- 0.1 \left(\frac{2a}{\ell}\right)^2\right]\,,\\
D_\text{coil}(\ell) &= \frac{2.22\mu_0 k_B T a}{\sqrt{l_p \ell}}\,.
\end{split}
\end{equation}
Here $2a$ is the diameter of the chain, $\mu_0 = 1/(6\pi \eta a)$ is
the Stokes mobility of a single sphere of radius $a$, and $\eta$ is
the viscosity of water.  $D_\text{cyl}(\ell)$ is the diffusion
constant of a stiff cylinder \cite{Tirado}, and $D_\text{coil}(\ell)$
gives the diffusion constant of an ideal chain in the flexible limit
where $\ell \gg l_p$.  We assume $\ell \ll l_p^3/a^2$, so
self-avoidance effects can be ignored~\cite{NetzAndelman}.  For
parameter values $a = 1$ nm and $l_p = 50$ nm, typical of DNA, these
effects are only important at scales greater than $\sim 10^2$ $\mu$m,
far larger than the chain lengths investigated in the experimental
studies discussed above.  Eq.~\eqref{eq:s2} is an interpolation
between the limiting cases given by $D_\text{cyl}(\ell)$ when $\ell
\ll l_p$ and $D_\text{coil}(\ell)$ when $\ell \gg l_p$.  In
Ref.~\citen{Schlagberger} the crossover exponent was determined to be
$\phi_2 =3$ and with the definition of $D_\text{cyl}(\ell)$ and
$D_\text{coil}(\ell)$ the constant was fixed at $C_2 =1$.  In the
present context we use $\phi_2$ and $C_2 \sim \text{O}(1)$ as
parameters that will be adjusted to fit the numerical data.

The asymptotic Zimm scaling is easily obtained from the expressions
written so far: In the flexible regime, for $\ell \gg l_p$, we have
$D_\text{coil} \sim (\ell l_p)^{-1/2}$ and from Eqs.~\eqref{eq:s1} and
~\eqref{eq:s1b} we find $\langle r^2_\text{intra} \rangle \sim \ell
l_p$.  Using $\langle r^2_\text{intra} \rangle \sim D t $ from
Eqs.~\eqref{diffusion} we thus obtain $\ell \sim t^{2/3} / l_p$ and
$\langle r^2_\text{intra} \rangle \sim t^{2/3}$, the well-known Zimm
scaling for flexible polymers~\cite{RubinsteinColby}.  In the stiff
polymer regime, for $\ell \ll l_p$, we have $D_\text{cyl} \sim
\ell^{-1} \log(\ell/a)$ and from Eq. \eqref{eq:s1b} we find $\langle
r^2_\text{intra} \rangle \sim \ell^3 / l_p $.  Again using $\langle
r^2_\text{intra} \rangle \sim D t $ we obtain this time $\ell \sim ( t
l_p \log t)^{1/4}$ and thus $\langle r^2_\text{intra} \rangle \sim
t^{3/4} l_p^{-1/4} \log^{3/4}t$.  The MSD scaling of semiflexible
polymers has pronounced logarithmic corrections in the stiff polymer
regime due to hydrodynamic effects.

To get an expression for the total MSD, $\langle r^2(t)\rangle$, that
includes the long-time regime, one must consider also the crossover
which occurs near time $t = \tau$, where $\ell(\tau) = L$, the total
contour length of the chain.  For $t \gg \tau$ the effective diffusion
constant is $D(L)$, and $\langle r^2(t) \rangle \approx 6 D(L) t$,
describing the trivial diffusion of the whole polymer coil.  $\tau$
corresponds approximately to the longest relaxation time of the
polymer.  This crossover is captured by yet another crossover
expression,
\begin{equation}\label{eq:s4}
\langle r^2(t) \rangle = \left[\langle r^2_\text{intra}(t)\rangle ^{\phi_3} +
  \{6D(L)t\}^{\phi_3} \right]^{1/\phi_3}\,,
\end{equation}
which gives the correct asymptotic scaling behavior for $\langle
r^2(t) \rangle$ in all time regimes.  In the results below, the
exponents $\phi_1$, $\phi_2$, $\phi_3$, and the three constants $C_1$,
$C_1^\prime$, $C_2$ in Eqs.~\eqref{eq:s1b}, \eqref{eq:s2}, and
\eqref{eq:s4} are chosen so that the heuristic scaling argument
approximately agrees, for long chain lengths, with the numerical
results of the MFT approach described in the next section.  The
best-fit values are: $\phi_1 = 0.84$, $\phi_2 = 3.15$, $\phi_3 =
3.62$, $C_1 = 3.65$, $C_1^\prime = 0.66$, $C_2 = 1.31$.

The full time dependence covering also the non-asymptotic behavior is
obtained by equating the expressions for $\langle r^2_\text{intra}(t)
\rangle$ in Eqs.~\eqref{eq:s1b} and \eqref{diffusion} and implicitly
solving for $\ell(t)$; we thus calculate $\langle r^2_\text{intra}(t)
\rangle$ as a function of $t$.  Plugging the result for $\langle
r^2_\text{intra}(t)\rangle$ into Eq.~\eqref{eq:s4} gives the total MSD
$\langle r^2(t)\rangle$.  The time evolution of $\langle
r^2(t)\rangle$ can be expressed through the effective exponent
$\alpha(t) = d \log \langle r^2(t) \rangle /d \log t$.
Fig.~\ref{fig:heur}(a) shows $\alpha(t)$ versus $t/\tau$ for chains of
total length $L = 12000a$, with various persistence lengths in the
range $l_p=25a - 4000a$.  The dot along each $\alpha(t)$ curve marks
$\alpha(\tau_p)$.  There is clearly an intermediate time regime,
within the range $\tau_p < t < \tau$, where $\alpha(t)$ dips below the
Zimm value of 2/3.  The minimum value of $\alpha(t)$ over all $t$
depends both on $L$ and $l_p$, as shown in Figs.~\ref{fig:heur}(b) and
(c), which plot $\min_t \alpha$ versus $l_p/L$ for several chain
lengths $L = 100a - 12000a$, and $\min_t \alpha$ versus $L/l_p$ for
several $l_p$ in the range $l_p = 50a - 1800a$. The overall variation
of $\min_t \alpha$ as a function of $L/a$ and $l_p/a$ is depicted in
the contour diagram of Fig.~\ref{fig:contours}.  The deviation from
Zimm behavior becomes more prominent with increasing $L$: the time
range where $\alpha < 2/3$ increases, and the values of $\min_t
\alpha$ decrease.  As seen in Fig.~\ref{fig:heur}(c), for fixed
$l_p/a$ the decrease in $\min_t \alpha$ with $L$ eventually saturates
for $L \gg l_p$.  The $\min_t \alpha$ curves in Fig.~\ref{fig:heur}(b)
all reach a minimum in the range $l_p/L \sim 0.01-0.04$.  The position
of the minimum decreases with $L$ approximately with the logarithmic
dependence $l_p/L \approx -0.013 + 0.26/\log(L/a)$.  The exponent
values at these minima, $\min_{l_p} \min_t \alpha$, also have a nearly
linear dependence on $1/\log(L/a)$, as can be seen in
Fig.~\ref{fig:heur}(d), where $\min_{l_p} \min_t \alpha$ goes from
0.677 at $L=200a$ to 0.591 at $L=12000a$.  The best-fit line is
$\min_{l_p} \min_t \alpha \approx 0.48+1.04/\log(L/a)$.  (If data from
$L$ much larger than the experimental range is also included, the $L =
\infty$ extrapolation of $\min_{l_p} \min_t \alpha$ shifts from
0.48, approaching 1/2.)  The growing deviation from Zimm behavior with
$L$ is possibly related to the observation in Ref.~\citen{Shusterman}
that the intermediate Rouse regime becomes more noticeable at longer
coil sizes, occupying a larger range of times.  However, in contrast
to Ref.~\citen{Shusterman}, the exponent $\alpha$ never reaches the
true Rouse value of $1/2$ even at the longest realistic chain lengths.
For $l_p = 50a$, corresponding to the DNA persistence length, $\min_t
\alpha$ at $L = 12000a$ is 0.602.

The origin of this intermediate regime where $\alpha(t) < 2/3$ can be
linked to the crossover behaviors of $\langle r^2_\text{intra}
\rangle$ and $D(\ell)$. Assume $\ell$ is in the range $l_p < \ell < L$
and is sufficiently large that $\langle r^2_\text{intra} \rangle \sim
\ell l_p$, but small enough that $D(\ell)$ has not reached the
asymptotic limit $D_\text{coil}(\ell) \sim (\ell l_p)^{-1/2}$.  Since the total MSD
$\langle r^2 \rangle \sim \langle r^2_\text{intra} \rangle$ in this
regime, one can use $\langle r^2_\text{intra} \rangle \sim D t$ and
$\langle r^2_\text{intra} \rangle \sim \ell l_p$ to relate the
effective exponent $\alpha = d\log \langle r^2 \rangle/d\log t$ to
$D(\ell)$, giving $\alpha = (1-(\ell/D)\partial D /\partial
\ell)^{-1}$.  As $D(\ell)$ is in the crossover region between
$D_\text{cyl}(\ell)$ and $D_\text{coil}(\ell)$, it must decrease with
$\ell$ slower than $\ell^{-1} \log(\ell/a)$, but faster than
$\ell^{-1/2}$.  These two limits mean that $\alpha$ is bounded by
$1/(2-\log^{-1}(\ell/a)) > 1/2$ from below, and $2/3$ from above,
corresponding precisely to the intermediate dynamical regime.

The existence of this regime will be confirmed through the Brownian
dynamics and MFT calculations described in the next two sections.  In
Sec.~\ref{results} we will see that the qualitative trends illustrated
in Fig.~\ref{fig:heur}(a)-(d) agree very well with the results from
the more sophisticated MFT approach and thus allow for a simple
explanation of sub-Zimm scaling behavior in terms of hydrodynamic
effects on the diffusion behavior of a semiflexible polymer in the
crossover between two limiting regimes.

\section{Brownian Dynamics Simulation}\label{brownian}

For the numerical Brownian dynamics simulations
\cite{Ermak,ErmakMcCammon} we model the polymer as a connected chain
of $M$ spheres, each having radius $a$ and position $\mb{r}_i(t)$,
$i=1,\ldots,M$.  The sphere positions evolve in time according to the
Langevin equation,
\begin{equation}\label{eq:n1}
\frac{d\mb{r}_i(t)}{dt} = \sum_{j=1}^M \overleftrightarrow{\bs{\mu}}_{ij} \cdot \left(-\frac{\partial U(\mb{r}_1,\ldots,\mb{r}_M)}{\partial \mb{r}_j} \right)+\bs{\xi}_i(t)\,,
\end{equation}
appropriate for the low Reynolds number regime.  Here
$\overleftrightarrow{\bs{\mu}}_{ij}$ is the Rotne-Prager tensor
\cite{RotnePrager} describing hydrodynamic interactions between the
monomers,
\begin{equation}\label{eq:n2}
\begin{split}
\overleftrightarrow{\bs{\mu}}_{ij} =& \mu_0 \delta_{i,j} \overleftrightarrow{\mb{1}} + (1-\delta_{i,j})\left(\frac{1}{8\pi\eta r_{ij}} \left[\overleftrightarrow{\mb{1}} +\frac{\mb{r}_{ij}\otimes\mb{r}_{ij}}{r_{ij}^2}\right]\right.\\
&\qquad\qquad\left. + \frac{a^2}{4\pi\eta r_{ij}^3}\left[ \frac{\overleftrightarrow{\mb{1}}}{3} - \frac{\mb{r}_{ij}\otimes\mb{r}_{ij}}{r_{ij}^2}\right]\right)\,,
\end{split}
\end{equation}
where $\mb{r}_{ij} \equiv \mb{r}_i
- \mb{r}_j$, and $\overleftrightarrow{\mb{1}}$ is the $3\times 3$
identity matrix.  The stochastic velocity $\bs{\xi}_i(t)$ in
Eq.~\eqref{eq:n1} is Gaussian, with correlations given by the
fluctuation-dissipation equation:
\begin{equation}\label{eq:n3}
\langle \bs{\xi}_i(t) \otimes \bs{\xi}_j(t^\prime) \rangle = 2 k_B T
\overleftrightarrow{\bs{\mu}}_{ij} \delta(t-t^\prime)\,.
\end{equation}
The final component of the model is the elastic potential
$U(\mb{r}_1,\ldots,\mb{r}_M)$ in Eq.~\eqref{eq:n1}, depending on the
positions of the spheres.  This potential consists of two parts,
\begin{equation}\label{eq:n4}
U = U_\text{WLC} + U_\text{LJ}\,,
\end{equation}
with
\begin{equation}\label{eq:n5}
\begin{split}
U_\text{WLC} &= \frac{\gamma}{4a} \sum_{i=1}^{M-1} \left(r_{i+1,i}-2a\right)^2+\frac{\epsilon}{2a}\sum_{i=2}^{M-1}(1-\cos\theta_i)\,,\\
U_\text{LJ} &= \omega \sum_{i < j} \Theta(2a-r_{ij})\left[\left(\frac{2a}{r_{ij}}\right)^{12}-2\left(\frac{2a}{r_{ij}}\right)^{6} +1\right]\,.
\end{split}
\end{equation}
Here $\theta_i$ is the angle between $\mb{r}_{i+1,i}$ and
$\mb{r}_{i,i-1}$.  The $U_\text{WLC}$ term describes the stretching
and bending forces associated with the extensible worm-like chain
model, with stretching modulus $\gamma$ and bending modulus
$\epsilon$.  The latter is related to the persistence length $l_p$ of
the polymer through $\epsilon = l_p k_B T$.  For all the simulations
the stretching modulus is set at $\gamma = 200 k_BT/a$, which is large
enough that the total contour length of the polymer stays
approximately constant.  The $U_\text{LJ}$ term is a truncated
Lennard-Jones interaction with strength $\omega = 3k_BT$.

To implement Eq.~\eqref{eq:n1} numerically, we discretize it with time
step $\tau$, and use non-dimensionalized variables, measuring lengths
in units of $a$, times in units of $a^2 /(k_B T \mu_0)$, and energies in
units of $k_BT$.  For a given contour length $L=2aM$ and persistence
length $l_p$, the results described below are based on averages taken
from $15-50$ independent runs, each with time step $\tau = 3 \times
10^{-4}\: a^2/(k_BT\mu_0)$ and lasting for $\sim 10^8 - 10^9$ steps.
The first $10^6$ steps of a run are not used for data collection, and
afterwards output data are collected every $10^3 - 10^4$ steps.

\section{Mean-field Model of Polymer Dynamics}\label{mft}

The derivation of the mean-field model for semiflexible polymers is
described below.  Readers not interested in the technical details may
skip this section, the main result of which is Eq.~\eqref{eq:t25} for
the end-monomer MSD in terms of several parameters: the diffusion
constant $D$, relaxation times $\tau_n$ and coefficients $\Delta_n$.
All of these parameters can be determined for a given $L$ and $l_p$ by
obtaining the normal modes and numerically evaluating the hydrodynamic
interaction matrix $H$ as outlined in
Eqs.~\eqref{eq:t18}-\eqref{eq:t22}.

The analytical model of the polymer is a continuous space curve
$\mb{r}(s)$ of total length $L$, with contour coordinate $s$ in the
range $-L/2 \le s \le L/2$.  The simplest expression for the elastic
energy $U$ of the chain, incorporating the effects of rigidity, is
that of Kratky and Porod \cite{KratkyPorod},
\begin{equation}\label{eq:t1}
U = \frac{\epsilon}{2} \int ds\,\left(\frac{\partial
  \mb{u}(s)}{\partial s} \right)^2,
\end{equation}
where $\epsilon$ is the bending modulus introduced above, $\epsilon =
l_p k_B T$, and the tangent vector $\mb{u} \equiv \partial
\mb{r}/\partial s$ is subject to the constraint $\mb{u}^2(s) = 1$ at
each $s$.  As in Sec.~\ref{scaling} we assume $L \ll l_p^3/a^2$, so we
ignore self-avoidance effects.  The associated free energy is $F =
-\beta^{-1} \log Z$, with $\beta^{-1} \equiv k_BT$ and the partition function
$Z$ given by the functional integral,
\begin{equation}\label{eq:t2}
Z = \int {\cal D}\mb{u} \,\prod_s \delta\left( \mb{u}^2(s) -1\right) e^{-\beta U}\,.  
\end{equation}
The delta function enforcing the constraint can be equivalently
written using an additional functional integral over a complex
auxiliary field $\lambda(s)$,
\begin{equation}\label{eq:t3}
\begin{split}
Z &= \int_{-i\infty}^{i\infty} {\cal D}\lambda \int {\cal D}\mb{u}\, e^{-\beta U - \beta \int ds\,\lambda(s)(\mb{u}^2(s)-1)}\\
&\equiv \int_{-i\infty}^{i\infty} {\cal D}\lambda\, e^{-\beta {\cal F}[\lambda]}\,,
\end{split}  
\end{equation}
where we introduce the functional ${\cal F}[\lambda]$, and ignore
any constants arising from the normalization of the integral.  Since
calculations with this partition function are generally intractable
due to the tangent vector constraint, we employ the mean-field theory
(MFT) approach developed by Ha and Thirumalai
\cite{HaThirumalai1,HaThirumalai2}, evaluating the functional integral
over $\lambda(s)$ using a stationary-phase approximation:
\begin{equation}\label{eq:t4}
Z = \int_{-i\infty}^{i\infty} {\cal D}\lambda\, e^{-\beta {\cal F}[\lambda]} \approx e^{-\beta {\cal F}[\lambda_{\text{cl}}]}\,.
\end{equation}
Here $\lambda_{\text{cl}}(s)$ is the path satisfying the stationary-phase
condition $\delta {\cal F}/\delta \lambda(s) = 0$, and we have
neglected higher-order correction terms.  The resulting MFT
free energy $F_{\text{MF}} \equiv {\cal F}[\lambda_{\text{cl}}]$ takes the
form~\cite{HaThirumalai1}:
\begin{equation}\label{eq:t5}
F_{\text{MF}} = - \beta^{-1} \log  \int {\cal D}\mb{u}\,e^{-\beta U_{\text{MF}}}\,,
\end{equation}
where
\begin{equation}\label{eq:t6}
\begin{split}
U_{\text{MF}} =& \frac{\epsilon}{2} \int ds\,\left(\frac{\partial
  \mb{u}(s)}{\partial s} \right)^2 +\nu \int
ds\,\mb{u}^2(s)\\
 & + \nu_0 \left(\mb{u}^2(L/2) + \mb{u}^2(-L/2)\right)\,,
\end{split}
\end{equation}
and the constants $\epsilon$, $\nu$, and $\nu_0$ are related by:
\begin{equation}\label{eq:t7}
\sqrt{\frac{\nu \epsilon}{2}} = \nu_0 = \frac{3}{4}k_B T\,.
\end{equation}
Comparing the results of this mean-field analytical model to those
of the simulation described in the last section, we note that the
simulation potential energy in Eqs.~\eqref{eq:n4}-\eqref{eq:n5}
contains an additional extensional term with large parameter $\gamma
\gg k_B T /a$.  Applying the mean-field approach to an extensible
worm-like chain leads to a value of the effective stretching moduli
$\nu$ and $\nu_0$ slightly modified from that of Eq.~\eqref{eq:t7},
with corrections of the order of $\epsilon/\Gamma$, where $\Gamma =
4a^2 \gamma$.  The details are described in Appendix A.  Assuming
$\epsilon$ and $\Gamma$ are fixed in the continuum limit, and $\Gamma
\gg \epsilon$, one can ignore the finite extensibility of the chain in
constructing the mean-field theory.

The MFT elastic energy of Eq.~\eqref{eq:t6} can be derived in several
alternative ways: it was first proposed by Lagowski, Noolandi, and
Nickel \cite{LNN} as a modification of the Harris-Hearst model
\cite{HarrisHearst} that corrected chain inhomogeneities due to end
fluctuations; it was later independently derived from the maximum
entropy principle by Winkler, Reineker, and Harnau~\cite{HWR1}.  The
main consequence of the approximation is that the local constraint
$\mb{u}^2(s) = 1$ is relaxed and replaced by the condition $\langle
\mb{u}^2(s) \rangle = 1$.  If the relationship between the bending
modulus $\epsilon$ and $l_p$ is redefined as $\epsilon = (3/2)l_p k_B
T$ in $U_{\text{MF}}$, the tangent vector correlation
function has the same form as in the Kratky-Porod chain,
\begin{equation}\label{eq:t8}
\langle \mb{u}(s) \cdot \mb{u}(s^\prime) \rangle = \exp(-|s^\prime - s|/l_p)\,,
\end{equation}
Related quantities like the mean squared end-to-end distance and
radius of gyration are also correctly reproduced by the MFT
approximation with this redefinition of $\epsilon$, and thus we will
use it for the remainder of the paper.  This applies only to the MFT
elastic energy of Eq.~\eqref{eq:t6}; in the simulation $U_\text{WLC}$
of Eq.~\eqref{eq:n4} $\epsilon$ retains its original definition.

In deriving the diffusion behavior of the polymer in solution, we
follow an approach similar to that of HWR \cite{HWR2}, who first
studied the dynamical characteristics of the MFT model given by
Eq.~\eqref{eq:t6} using a hydrodynamic pre-averaging approximation
along the lines of the Zimm model \cite{Zimm,DoiEdwards}.  To describe
the time evolution of the chain $\mb{r}(s,t)$ in the presence of
hydrodynamic interactions, we start with the Langevin equation:
\begin{equation}\label{eq:t9}
\begin{split}
&\frac{\partial}{\partial t}\mb{r}(s,t) = -\int_{-L/2}^{L/2} ds^\prime\, \overleftrightarrow{\bs{\mu}}\left(s,s^\prime;\mb{r}(s,t)-\mb{r}(s^\prime,t)\right)\\
&\qquad\qquad\qquad\qquad\cdot \frac{\delta U_{\text{MF}}}{\delta \mb{r}(s^\prime,t)}  + \bs{\xi}(s,t)\,.
\end{split}
\end{equation}
Here the $\bs{\xi}(s,t)$ is the stochastic contribution, and
$\overleftrightarrow{\bs{\mu}}(s,s^\prime;\mb{x})$ is the continuum
version of the Rotne-Prager tensor in Eq.~\eqref{eq:n2} \cite{HWR2},
\begin{equation}\label{eq:t10}
\begin{split}
\overleftrightarrow{\bs{\mu}}(s,s^\prime;\mb{x}) =& \quad 2a\mu_0 \delta(s-s^\prime)\overleftrightarrow{\mb{1}}\\
&+ \Theta(x-2a)\left(\frac{1}{8\pi\eta x} \left[\overleftrightarrow{\mb{1}} +\frac{\mb{x}\otimes\mb{x}}{x^2}\right]\right.\\
&\left. + \frac{a^2}{4\pi\eta x^3}\left[ \frac{\overleftrightarrow{\mb{1}}}{3} - \frac{\mb{x}\otimes\mb{x}}{x^2}\right]\right)\,,
\end{split}
\end{equation}
with the $\Theta$ function excluding unphysical configurations.

The pre-averaging approximation consists of replacing
$\overleftrightarrow{\bs{\mu}}(s,s^\prime;\mb{r}(s,t)-\mb{r}(s^\prime,t))$ in
Eq.~\eqref{eq:t9}, which involves a complicated dependence on the
specific chain configuration at time $t$, with an average over all
equilibrium configurations,
$\overleftrightarrow{\bs{\mu}}_\text{avg}(s,s^\prime)$, that depends
only on the contour coordinates $s$ and $s^\prime$.  This tensor
$\overleftrightarrow{\bs{\mu}}_\text{avg}$ is defined as:
\begin{equation}\label{eq:t11}
\begin{split}
&\overleftrightarrow{\bs{\mu}}_\text{avg}(s,s^\prime) = \int d^3\mb{x}\, \overleftrightarrow{\bs{\mu}}(s,s^\prime;\mb{x}) G(s,s^\prime;\mb{x})\,,
\end{split}
\end{equation}
where $ G(s,s^\prime;\mb{x})$ is the equilibrium probability of finding two points at $s$ and $s^\prime$ along the polymer contour whose spatial
positions differ by the vector $\mb{x}$.  For the MFT model of
Eq.~\eqref{eq:t6}, this probability is \cite{HWR1}:
\begin{equation}\label{eq:t12}
G(s,s^\prime;\mb{x}) = \left( \frac{3}{2\pi \sigma(|s-s^\prime|)}\right)^{3/2} \exp\left(-\frac{3 \mb{x}^2}{ 2 \sigma(|s-s^\prime|)}\right)\,, 
\end{equation}
where $\sigma(l) \equiv 2l_p l - 2l_p^2
(1-\exp(-l/l_p))$, the mean squared end-to-end distance of a
chain of length $l$.  Plugging Eq.~\eqref{eq:t12} into
Eq.~\eqref{eq:t11} we find:
\begin{equation}\label{eq:t13}
\begin{split}
\overleftrightarrow{\bs{\mu}}_\text{avg}(s,s^\prime) = &\Biggl[ 2a\mu_0 \delta(s-s^\prime) +\frac{\Theta(|s-s^\prime|-2a)}{\eta\sqrt{6\pi^3 \sigma(|s-s^\prime|)}}\\
&\quad\cdot \exp\left(-\frac{6a^2}{\sigma(|s-s^\prime|)}\right)\Biggr]\overleftrightarrow{\mb{1}} \\
\equiv& \mu_\text{avg}(s-s^\prime) \overleftrightarrow{\mb{1}}\,.  
\end{split}
\end{equation}
For the same reason as in Eq.~\eqref{eq:t10}, we have added a
$\Theta$ function to the final result.

The pre-averaged version of the Langevin equation is thus
\begin{equation}\label{eq:t14}
\begin{split}
&\frac{\partial}{\partial t}\mb{r}(s,t) =\\
&\quad \int_{-L/2}^{L/2} ds^\prime\,\mu_\text{avg}(s-s^\prime)\left( -\frac{\delta U_{\text{MF}}}{\delta \mb{r}(s^\prime,t)}\right) + \bs{\xi}(s,t)\,.
\end{split}
\end{equation}
We assume the $\bs{\xi}(s,t)$ are Gaussian random vectors, whose
components $\xi^{(i)}(s,t)$ have correlations given by the
fluctuation-dissipation theorem:
\begin{equation}\label{eq:t14b}
\langle \xi^{(i)}(s,t) \xi^{(j)}(s^\prime,t^\prime)\rangle = 2k_B T\delta_{ij}\delta(t-t^\prime)\mu_\text{avg}(s-s^\prime).
\end{equation}
Using $U_{\text{MF}}$ from Eq.~\eqref{eq:t6}, the force term in
Eq.~\eqref{eq:t14} can be written as
\begin{equation}\label{eq:t15}
-\frac{\delta U_{\text{MF}}}{\delta \mb{r}(s^\prime,t)} =-\epsilon \frac{\partial^4}{\partial s^4}\mb{r}(s^\prime,t)+2\nu\frac{\partial^2}{\partial s^2} \mb{r}(s^\prime,t)\,, 
\end{equation}
with free-end boundary conditions at $s = \pm L/2$ of
the form,
\begin{equation}\label{eq:t16}
\begin{split}
\epsilon \frac{\partial^3}{\partial s^3}\mb{r}(\pm L/2,t) - 2\nu \frac{\partial}{\partial s} \mb{r}(\pm L/2,t) &= 0\,,\\
\mp\epsilon \frac{\partial^2}{\partial s^2}\mb{r}(\pm L/2,t) - 2\nu_0 \frac{\partial}{\partial s} \mb{r}(\pm L/2,t) &= 0\,.
\end{split}
\end{equation}
To rewrite the Langevin equation in matrix form, we assume
$\bs{\xi}(s,t)$ satisfies similar boundary conditions to
$\mb{r}(s,t)$, and expand both $\mb{r}(s,t)$ and $\bs{\xi}(s,t)$ in
normal modes $\psi_n(s)$, with amplitudes $\mb{p}_n(t)$ and $\mb{q}_n(t)$
respectively:
\begin{equation}\label{eq:t17}
\mb{r}(s,t) = \sum_{n=0}^\infty \mb{p}_n(t) \psi_n(s), \quad \bs{\xi}(s,t) = \sum_{n=0}^\infty \mb{q}_n(t) \psi_n(s)\,.
\end{equation}
We choose the normal modes $\psi_n(s)$ to be eigenfunctions of the
differential operator in Eq.~\eqref{eq:t15}, satisfying
\begin{equation}\label{eq:t18}
\epsilon \frac{\partial^4}{\partial s^4}\psi_n(s)-2\nu\frac{\partial^2}{\partial s^2} \psi_n(s)=\lambda_n \psi_n(s)\,, 
\end{equation}
for eigenvalues $\lambda_n$.  These $\psi_n(s)$ take the form \cite{HWR2}:
\begin{equation}\label{eq:t19}
\begin{split}
\psi_0(s) &= \sqrt{\frac{1}{L}}\,,\\
\psi_n(s) &= \sqrt{\frac{A_n}{L}}\left( \alpha_n \frac{\sin \alpha_n s}{\cos \alpha_n L/2} \right. \\
&\qquad \left. + \beta_n \frac{\sinh \beta_n s}{\cosh \beta_n L/2}\right)\,, \quad n\: \text{odd},\\
\psi_n(s) &= \sqrt{\frac{A_n}{L}}\left( -\alpha_n \frac{\cos \alpha_n s}{\sin \alpha_n L/2} \right. \\
&\qquad \left. + \beta_n \frac{\cosh \beta_n s}{\sinh \beta_n L/2}\right)\,, \quad n\: \text{even},
\end{split}
\end{equation}
with
\begin{equation}\label{eq:t20}
\beta^2_n - \alpha^2_n = 2\nu/\epsilon\,, \quad \lambda_0=0\,,\quad \lambda_n = \epsilon \alpha_n^4 + 2\nu \alpha_n^2\,.
\end{equation} 
The constants $\alpha_n$ and $\beta_n$ can be determined from the
boundary conditions in Eq.~\eqref{eq:t16}, while the $A_n$ are
normalization coefficients.  Using Eqs.~\eqref{eq:t17},
\eqref{eq:t18}, and the orthonormality of the $\psi_n$,
Eqs.~\eqref{eq:t14} and \eqref{eq:t14b} become:
\begin{equation}\label{eq:t21}
\begin{split}
\frac{\partial}{\partial t}\mb{p}_n(t) &= - \sum_{m=0}^\infty H_{nm} \lambda_m \mb{p}_m(t) + \mb{q}_n(t)\,,\\
 \langle q_{ni}(t) q_{mj}(t^\prime) \rangle &= 2k_B T\delta_{ij}\delta(t-t^\prime)H_{nm}\,,
\end{split}
\end{equation}
where
\begin{equation}\label{eq:t22}
H_{nm} = \int_{-L/2}^{L/2}ds \int_{-L/2}^{L/2}ds^\prime\,\psi_n(s)\mu_\text{avg}(s-s^\prime)\psi_m(s^\prime)\,.
\end{equation}
The matrix elements $H_{nm}$ can be evaluated through numerical
integration.  HWR neglect the off-diagonal portion of this interaction
matrix $H$, since the diagonal elements typically dominate.  However,
as we will show later, this approximation leads to an inaccurate
description of the simulation results for the end-monomer dynamics at
short times, demonstrating that the off-diagonal elements are
negligible only at times longer than the bending relaxation time.  A
more accurate approach is to take the whole matrix $H$, keep only the
leading $N \times N$ sub-block (describing the interactions among the
$N$ slowest-relaxing modes), and exactly solve the resulting
finite-dimensional version of Eq.~\eqref{eq:t21}.  An appropriate
value for $N$ can be estimated as follows.  For the oscillation
described by mode $\psi_n(s)$ from Eq.~\eqref{eq:t19}, the distance
between successive nodes is approximately $L/n$.  The high-frequency
cutoff of this distance is on the order of two monomer diameters $4a$,
so that only modes with $n \lesssim L/4a$ should be considered.  Thus
the natural choice is $N = L/4a$.  In the results described in
Sec.~\ref{results}, we use this choice for all chains with $L \le
1600a$.  For longer chains with $L > 1600a$, calculation of the full
matrix becomes numerically unfeasible due to roundoff errors in the
highly oscillatory integrals of Eq.~\eqref{eq:t22}.  Thus for these
chains we truncate $N$ at the maximum value of $N=400$.  This approach
gives accurate results at time scales much larger than the relaxation
time of the $n=400$ mode, which is always the case for the time ranges
of interest.

To implement this approach, let $J$ be the $N \times N$ matrix with
elements $J_{nm} = H_{nm} \lambda_m$, $\Lambda_n$ the eigenvalues of
$J$, and $C$ the matrix diagonalizing $J$: $(C J C^{-1})_{nm} =
\Lambda_n \delta_{nm}$.  Assuming the eigenvalues $\Lambda_n$ are
distinct, and using the fact that $H$ is symmetric, it can also be
shown that the matrix $C$ diagonalizes $H$ through the congruent
transformation: $(C H C^T)_{nm} = \Theta_n \delta_{nm}$, defining
diagonal elements $\Theta_n$ \cite{Zimm}.  If we introduce a new set
of orthogonal functions $\Psi_n(s)$ and the associated amplitudes
$\mb{P}_n(t)$, $\mb{Q}_n(t)$,
\begin{equation}\label{eq:t23}
\begin{split}
\Psi_n (s) &= \sum_{m=0}^{N-1} \psi_m(s) \left(C^{-1}\right)_{mn}\,,\\
\mb{P}_n(t) &= \sum_{m=0}^{N-1} C_{nm} \mb{p}_m(t)\,,\quad \mb{Q}_n(t) = \sum_{m=0}^{N-1} C_{nm} \mb{q}_m(t)\,,
\end{split}
\end{equation} 
then Eq.~\eqref{eq:t21} becomes
\begin{equation}\label{eq:t24}
\begin{split}
\frac{\partial}{\partial t}\mb{P}_n(t) &= - \Lambda_n \mb{P}_n(t) + \mb{Q}_n(t)\,,\\
 \langle Q_{ni}(t) Q_{mj}(t^\prime) \rangle &= 2k_B T\delta_{ij}\delta(t-t^\prime)\Theta_{n}\delta_{nm}\,.
\end{split}
\end{equation}
This equation can be solved directly to yield the end-monomer MSD:
\begin{equation}\label{eq:t25}
\begin{split}
\langle r^2(t) \rangle &\equiv \langle
  \left(\mb{r}(L/2,t)-\mb{r}(L/2,0)\right)^2 \rangle\\&= 6 D t
  + \sum_{n=1}^{N-1} \Delta_n (1-e^{-t/\tau_n})\,,
\end{split}
\end{equation}
where the diffusion constant $D = k_B T \Theta_0 \Psi_0^2(L/2)$, the
relaxation times $\tau_n = \Lambda_n^{-1}$, and $\Delta_n = 6 k_B T \tau_n
\Theta_n \Psi_n^2(L/2)$.

\section{Results}\label{results}

\begin{figure}
\begin{center}
\includegraphics*{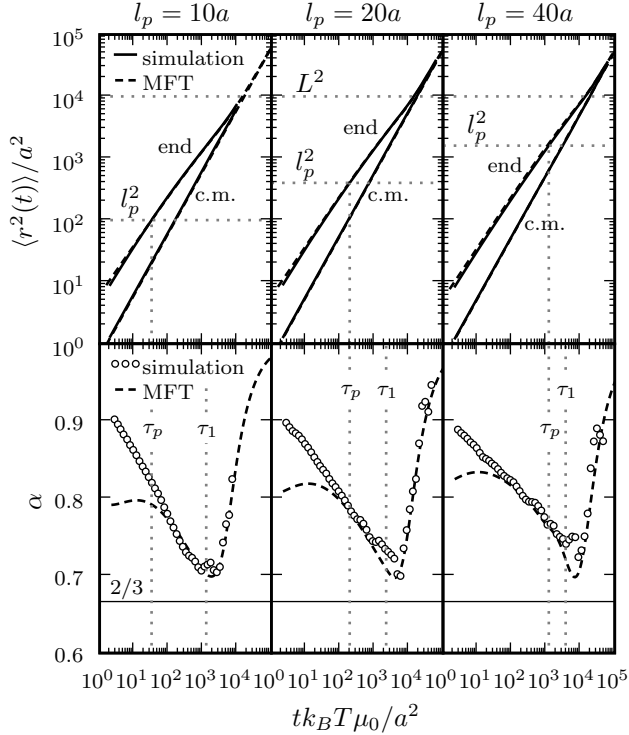}
\end{center}
\caption{The top panels show simulation and MFT results for the
  end-monomer and center-of-mass MSD $\langle r^2(t) \rangle/a^2$ for
  chains of length $L = 100a$ and persistence lengths $l_p = 10a$,
  $20a$, and $40a$ ($a =$ monomer radius).  The bottom panels show the
  local slope $\alpha = d \log \langle r^2(t) \rangle /d \log t$ for
  the end-monomer MSD, with a horizontal line at $2/3$ marking the
  Zimm theory prediction.  Two times are indicated by dotted vertical
  lines: $\tau_p$, the time at which the end-monomer $\langle
  r^2(\tau_p) \rangle = l_p^2$; and $\tau_1$, the longest relaxation
  time of the polymer, as determined from the numerical simulations.}
\label{fig:M50msd}
\end{figure}

\begin{figure}
\begin{center}
\includegraphics*{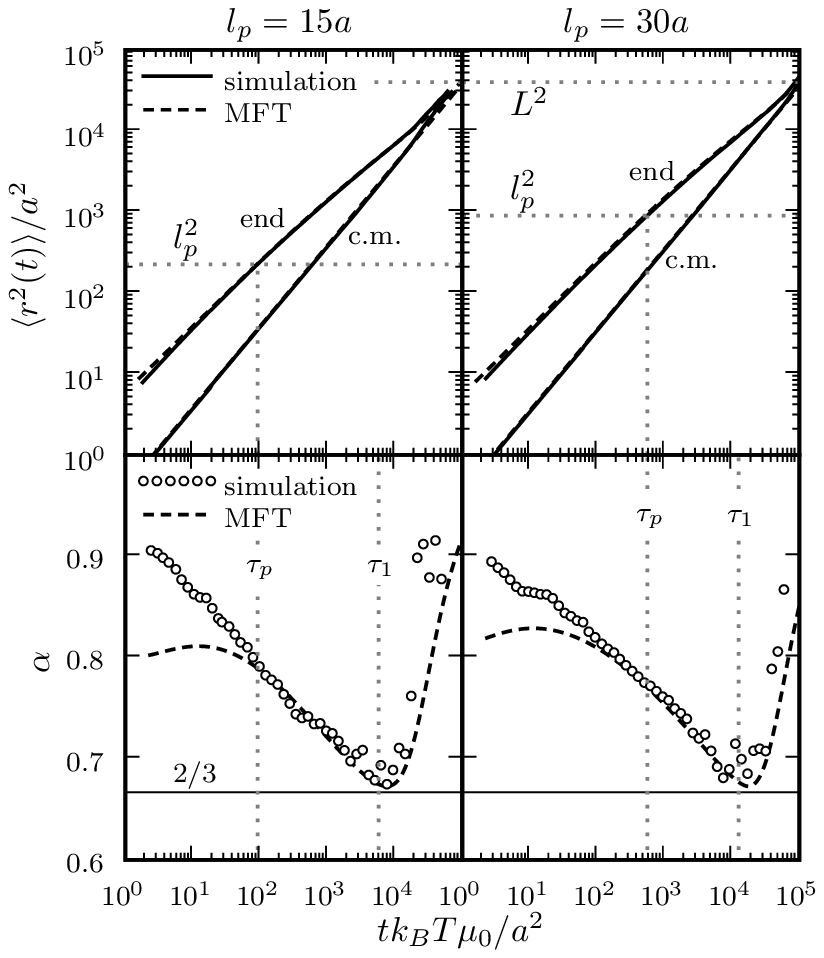}
\end{center}
\caption{Same as in Fig.~\ref{fig:M50msd}, but for chains of length $L
  = 200a$ and persistence lengths $l_p = 15a$ and $30a$.}
\label{fig:M100msd}
\end{figure}

\begin{figure}
\begin{center}
\includegraphics*{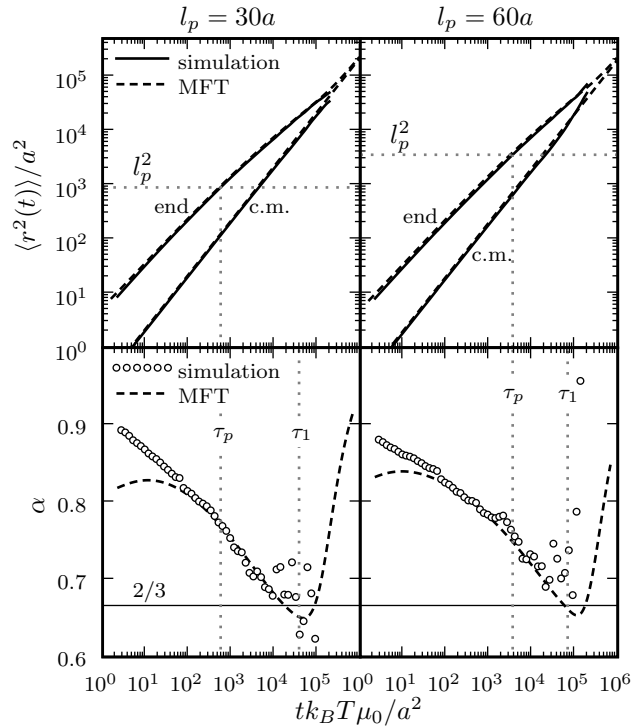}
\end{center}
\caption{Same as in Fig.~\ref{fig:M50msd}, but for chains of length $L
  = 400a$ and persistence lengths $l_p = 30a$ and $60a$.}
\label{fig:M200msd}
\end{figure}

The Brownian dynamics simulation and MFT results for the end-monomer
and center-of-mass MSD are shown in
Figs. \ref{fig:M50msd}-\ref{fig:M200msd} for chain lengths of
$L=100a$, $200a$, and $400a$ respectively, at various persistence
lengths $l_p$.  We also show in the bottom panels of each figure the
effective local exponent $\alpha(t) = d \log \langle r^2(t) \rangle /
d \log t$ of the end-monomer MSD curve.

We find in both the simulation and MFT results that $\alpha(t)$ passes
through a minimum in the intermediate time range where $l_p^2 <
\langle r^2(t) \rangle < r^2_\text{ee}(L)$.  The location of this
minimum is on the order of $\tau_1$, the longest relaxation time of
the polymer.  For $t \gg \tau_1$, as the end-monomer curve approaches
the center-of-mass MSD, $\langle r^2_{\text{c.m.}}(t) \rangle = 6 D
t$, the local slope $\alpha(t)$ tends toward the limiting value of 1.
On the other hand, for $t < \tau_p$, where $\langle r^2(t) \rangle <
l_p^2$, the stiffness of the polymer dominates, and $\alpha(t)$ varies
in the range $\approx 0.8 - 0.9$.  We will discuss both the
intermediate and the short-time regimes in more detail below.

\begin{figure}
\begin{center}
\includegraphics*{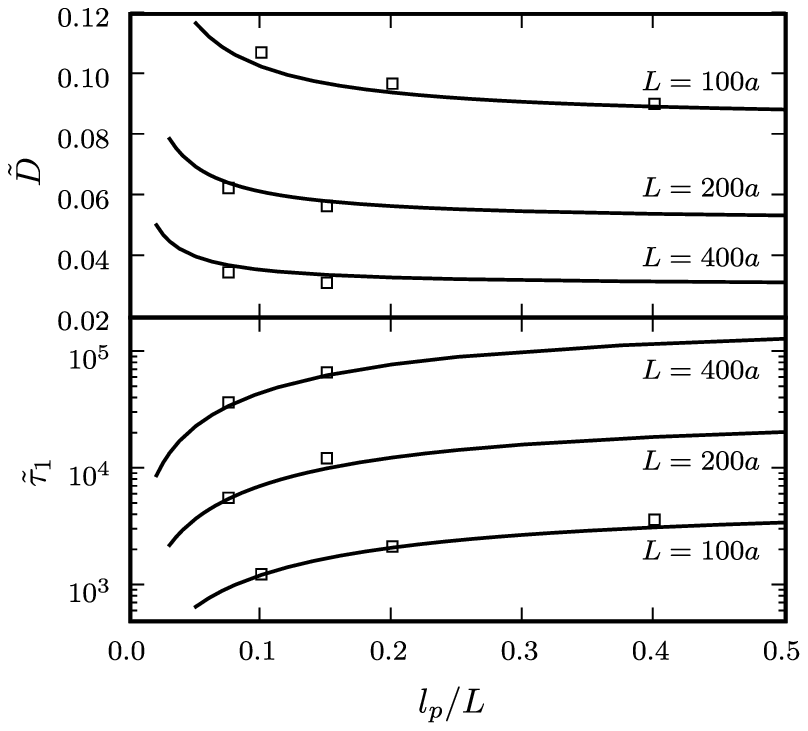}
\end{center}
\caption{Diffusion constant $\tilde{D} = D/(k_B T \mu_0)$ (top) and
  longest relaxation time $\tilde{\tau}_1 = \tau_1 k_B T \mu_0 /a^2$
  (bottom) as a function of $l_p/L$ for three different chain lengths
  $L$.  The squares are Brownian dynamics simulation results, while
  the solid curves are calculated from MFT.}
\label{fig:diff_tau}
\end{figure}

\begin{figure*}
\begin{center}
\includegraphics*{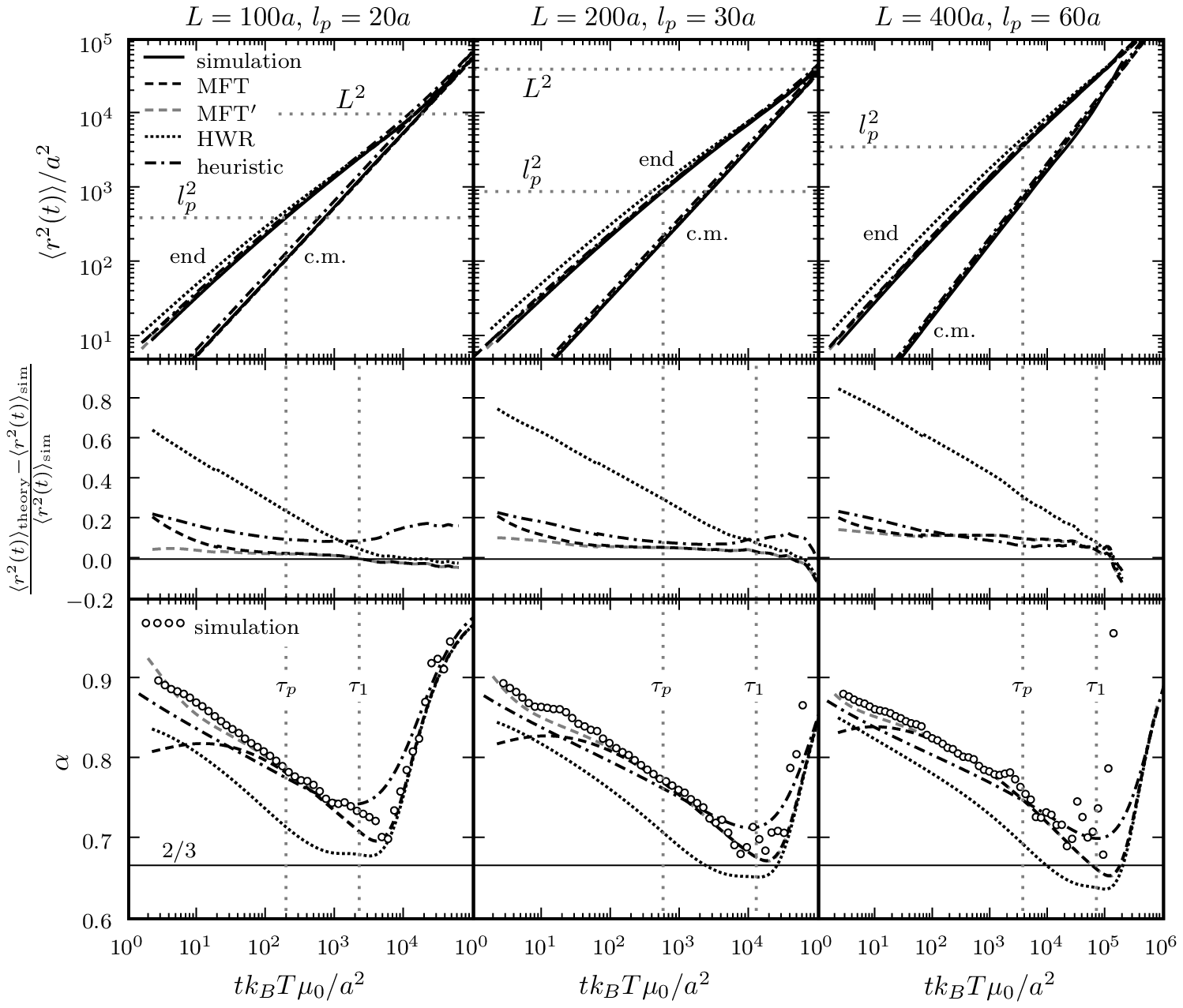}
\end{center}
\caption{Comparison of the simulation results for three semiflexible
  chains with several analytical approaches: the heuristic scaling
  argument described in Sec.~\ref{scaling}, the MFT described in
  Sec.~\ref{mft}, and the MFT$^\prime$, HWR \cite{HWR2} models
  described in Sec.~\ref{results}.  The top panels show the
  end-monomer and center-of-mass MSD $\langle r^2(t) \rangle$; the
  middle panels show the relative difference between the various
  theoretical results for the end-monomer MSD and the simulation data;
  the bottom panels show the local slope $\alpha = d\log \langle
  r^2(t) \rangle /d \log t$ for the end-monomer MSD.  Two times are
  indicated by dotted vertical lines: $\tau_p$, the time at which the
  end-monomer $\langle r^2(\tau_p) \rangle = l_p^2$; and $\tau_1$, the
  longest relaxation time of the polymer.  The $\tau_1$ and $\tau_p$
  values shown are from the numerical simulations.}
\label{fig:theory_comp}
\end{figure*}

There is very good agreement between MFT end-monomer MSD predictions
and simulation results in time regimes where simulation results have
sufficiently converged to make a comparison.  (For $\langle r^2(t)
\rangle$ this comparison is possible for nearly the whole simulation
time range; for $\alpha(t)$ the numerical uncertainty at the largest times
becomes significant, and is on the order of the scatter in the plotted
data points.)  Additionally, dynamical parameters like the diffusion
constant $D$ and relaxation time $\tau_1$ determined from the
simulation data compare favorably with their MFT values, as shown in
Fig.~\ref{fig:diff_tau}.  The values of $D$ were obtained from the
simulations by fitting the center-of-mass MSD data to the
straight-line form $\langle r^2_\text{c.m.}(t)\rangle = 6 D t$.  To
extract $\tau_1$, the autocorrelation function of the end-to-end
vector was calculated, $C_\text{ee}(t) = \langle \mb{r}_\text{ee}(t)
\cdot \mb{r}_\text{ee}(0) \rangle$, where $\mb{r}_\text{ee}(t) =
\mb{r}_M(t) - \mb{r}_1(t)$.  For sufficiently large $t$, this function
takes the form of a simple exponential decay, $C_\text{ee}(t) \sim
\exp(-t/\tau_1)$, from which $\tau_1$ can be estimated.  For both $D$
and $\tau_1$, only data points for which convergence was achieved were
included in the fitting (the criterion for convergence was that the
local slope $d\log \langle \mb{r}^2_\text{c.m.}(t)\rangle /d\log t
\approx 1$.)

The main discrepancies between the two approaches are in the local
slopes of the MSD curves at the shortest times, $t \lesssim 10^2\:
a^2/(k_BT\mu_0)$.  This can be explained by the fact that the
small-scale motions at short times are particularly sensitive to the
discrete nature of the polymer chain and the more strongly fixed
monomer-monomer separation in the simulation, thus giving rise to
differences with the continuum mean-field approximation.  In fact we
can make the MFT mimic the simulation more closely if we exclude the
contributions of a fraction of the highest modes in the sum of
Eq.~\eqref{eq:t25}, by changing the upper limit from $N-1$ to
$N^\prime -1$, where $N^\prime = c L/4a$, $0 < c < 1$.  A value of $c
\approx 1/2$ gives the closest approximation to the simulation MSD and
$\alpha(t)$ curves, irrespective of $L$ and $l_p$.  This roughly
corresponds to excluding modes where the distance between nodes is
shorter than four monomer diameters.  The results are shown in
Fig.~\ref{fig:theory_comp} for three different chains, with the
modified MFT labeled as MFT$^\prime$.  The long-time behavior is
unaffected by removing the highest modes, but at short times the
MFT$^\prime$ $\alpha(t)$ curves fit the simulation data much more
closely.  We rationalize this as being due to an effective cutoff of
fluctuations at small wavelengths due to the spring stiffness in the
simulation, which is not represented well by the Gaussian MFT elastic
energy in Eq.~\eqref{eq:t6}.

\begin{figure}
\begin{center}
\includegraphics*{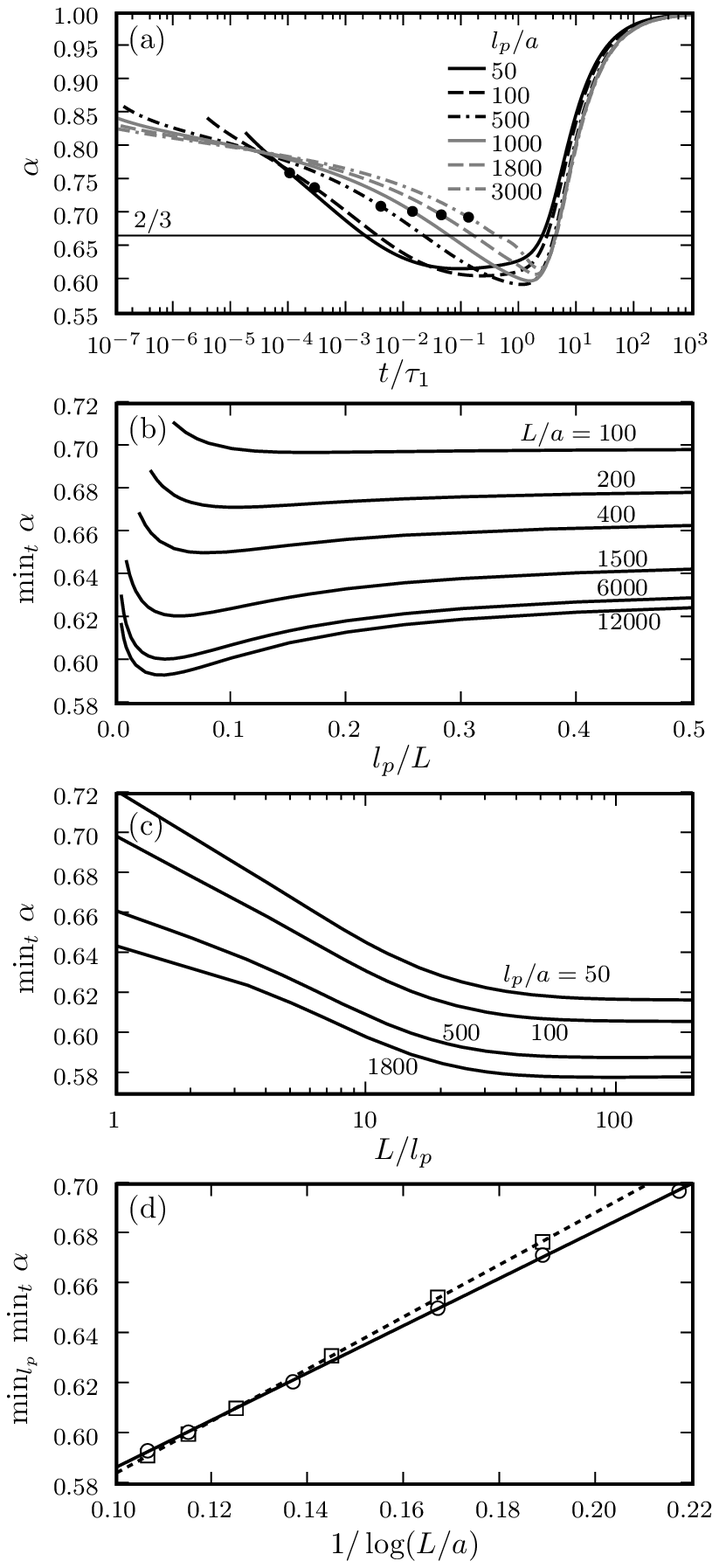}
\end{center}
\caption{Results of the MFT for various semiflexible chains.  (a) The
  end-monomer MSD effective exponent $\alpha = d \log \langle r^2(t)
  \rangle /d \log t$ versus $t/\tau_1$ for polymers of diameter $2a$,
  contour length $L = 12000a$, and various persistence lengths $l_p =
  50a - 3000a$.  The dot along each $\alpha(t)$ curve marks
  $\alpha(\tau_p)$.  The time $\tau_p$ is where $\langle r^2 (\tau_p)
  \rangle = l_p^2$, and $\tau_1$ is the longest relaxation time of the
  polymer.  (b) The minimum value of $\alpha$ over all $t$ versus
  $l_p/L$ for several chain lengths $L = 100a - 12000a$.  (c) The
  minimum value of $\alpha$ over all $t$ versus $L/l_p$ for several
  $l_p = 50a-1800a$.  (d) The circles show the minimum value of
  $\alpha$ in each of the curves in panel (b), plotted as a function
  of $1/\log(L/a)$.  Superimposed is a straight-line fit to the data
  points.  The squares, with the dashed straight-line fit, are the
  corresponding results of the heuristic scaling argument, taken from
  Fig.~\ref{fig:heur}(d).}
\label{fig:mft_results}
\end{figure}

\begin{figure}
\begin{center}
\includegraphics*[width=\textwidth]{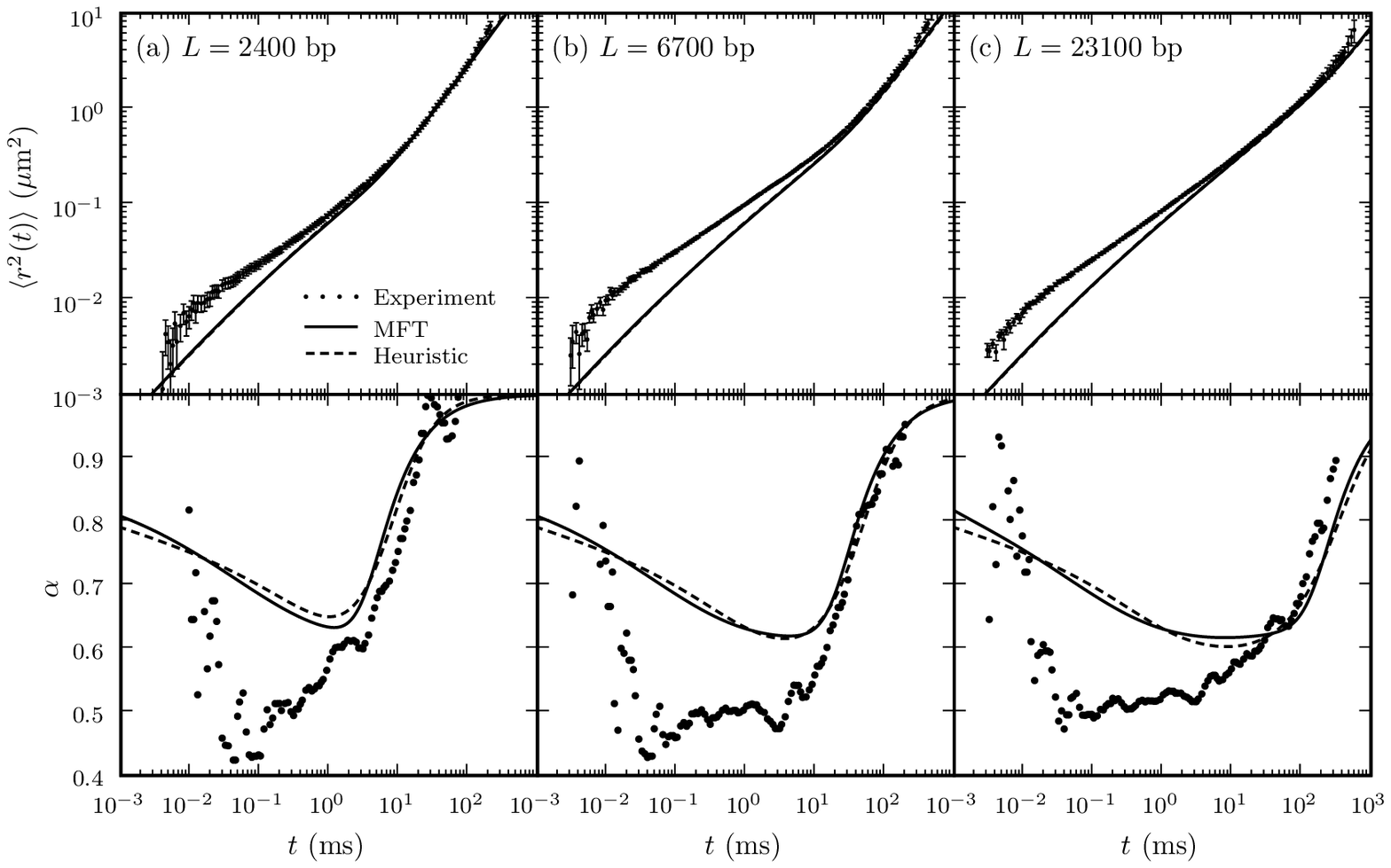}
\end{center}
\caption{Comparison of the MFT and heuristic scaling results to
  experimental MSD data taken from Ref.~\citen{Shusterman} for three
  lengths of double-stranded DNA: (a) 2400 bp; (b) 6700 bp; (c) 23100
  bp.  The dimensional parameters in the theories are fixed at: $a =
  1$ nm, rise per bp $= 0.34$ nm, $l_p = 50$ nm, $T = 293$K, viscosity
  of water at 293K = 1 mPa s.  The top panels show the end-monomer MSD
  $\langle r^2(t) \rangle$, and the bottom panels show the
  corresponding effective local exponent $\alpha = d\log \langle r^2(t) \rangle
  /d \log t$. }
\label{fig:exp_comp}
\end{figure}

In Fig.~\ref{fig:theory_comp} we also show the MSD and $\alpha(t)$
curves calculated using the HWR model.  This model follows the same
basic approach as in Sec.~\ref{mft}, with two additional
approximations: (i) only the diagonal elements $H_{nn}$ of
Eq.~\eqref{eq:t22} are used; (ii) for $n > 1$, the $H_{nn}$ are
evaluated approximately as \cite{HWR2}
\begin{equation}\label{eq:r4}
H_{nn} \approx 2\sqrt{\frac{6}{\pi}} \frac{\mu_0 a}{L} \int_d^L ds\,
\frac{L-s}{\sqrt{\sigma(s)}} \exp\left(-\frac{3d^2}{2\sigma(s)} \right)\cos \alpha_{n} s\,.
\end{equation}
The net effect of these approximations is negligible only for $t
\gtrsim \tau_1$, where the HWR and MFT results overlap.  For $t <
\tau_1$ there are significant differences with respect to the
simulations.  Here the HWR model overestimates the end-monomer MSD and
underestimates $\alpha(t)$.  The discrepancy is only slightly reduced
by avoiding the approximation of Eq.~\eqref{eq:r4}; the main weakness
of the HWR model is that the off-diagonal matrix elements $H_{nm}$ are
not included in the calculation.  Taking these into account, as the
MFT results demonstrate, gives a much more accurate description of the
simulation data at short and intermediate times.  Despite these
differences, the sub-Zimm scaling regime exists in the HWR model, and
the deviation below 2/3 is even larger than in the MFT.  (The sub-Zimm
scaling is also implicitly evident in related quantities calculated
from the HWR approach, like the fluorescence correlation function
studied in Ref.~\citen{Winkler}.)

For comparison, Fig.~\ref{fig:theory_comp} also shows the $\langle
r^2(t)\rangle$ and $\alpha(t)$ curves calculated from the heuristic
scaling argument of Sec.~\ref{scaling}.  Despite its simplicity, it is
able to capture the trends of the simulation and MFT data quite well,
though for shorter chain lengths it gives a shallower dip in $\alpha$
within the intermediate regime.

Given the success of the MFT at reproducing the simulation results, it
is interesting to see what the theory predicts for longer chain
lengths where Brownian hydrodynamics simulations become impractical.
Fig.~\ref{fig:mft_results}(a) shows $\alpha(t)$ curves for $L =
12000a$, $l_p = 50a - 3000a$, with the point $\alpha(\tau_p)$ on each
curve marked by a dot.  As in the shorter chains, there is a broad dip
in $\alpha(t)$ between $\tau_p$ and $\tau_1$, but the minimum of
$\alpha(t)$ has been shifted to below $2/3$.  In fact the dependence
of this minimum on $L$ and $l_p$, illustrated in
Fig.~\ref{fig:mft_results}(b-d), is qualitatively the same as
that derived from the heuristic scaling argument in
Fig.~\ref{fig:heur}(b-d): there is a general trend of $\min_t \alpha$
decreasing with $L$, and in particular the smallest value possible at
a given $L$, $\min_{l_p} \min_t \alpha$, has a nearly linear
dependence on $1/\log(L/a)$ (the heuristic result from
Fig.~\ref{fig:heur}(d) is also drawn for reference).

At $l_p = 50a$, corresponding to the persistence length of DNA,
$\min_t \alpha$ ranges from 0.698 at $L=100a$ to 0.617 at $L=12000a$.
Within this range we can make a detailed comparison for three
particular chain lengths where experimental MSD data for
double-stranded DNA is available from Ref.~\citen{Shusterman}: $L =
816a$, $2278a$, and $7854a$, or equivalently 2400 bp, 6700 bp, and
23100 bp (using $a=1$ nm, and a rise per base pair of 0.34 nm).  The
experimental end-monomer MSD for these three cases is shown in the top
panels of Fig.~\ref{fig:exp_comp}.  The bottom panels show the local
slope $\alpha(t)$, which can be estimated at each $t$ by fitting
straight lines to the log-log plot of MSD data points with times $t_i$
within a small range around $t$, defined by the condition $|\log_{10}
t_i/t| < 0.15$.  Together with the experimental results for the MSD
and $\alpha(t)$ are the curves predicted by the MFT and heuristic
scaling argument.  Besides the length scale parameters mentioned
above, the other dimensional variables in the system are set at the
following values (taken from the literature and the experimental
conditions): $l_p = 50$ nm, $T = 293$K, viscosity of water at 293K = 1
mPa s.  At longer times ($> 10$ ms) there is quite good agreement
between both theories and experimental data, particularly notable
since there is no fitting parameter involved in the MFT.  The
discrepancies arise in the short and intermediate time regime, where
the experimental MSD is consistently higher than the theoretical one,
the difference increasing to roughly a factor of $2-3$ at the shortest
times measured.  The discrepancy in the MSD is on the order of
$0.001-0.01$ $\mu\text{m}^2$, corresponding to length scales roughly
$30-100$ nm.  Experimental data seems to indicate faster displacement
of the end monomer at very short times followed by slower increase
of the MSD at intermediate times compared to the theoretical
predictions.

The effect of the higher experimental MSD is to push the local slope
down relative to the theoretical value.  Thus the intermediate
dynamical regime, in the range $0.01 - 10$ ms, is characterized by a
broad region with $\alpha(t)$ close to 0.5, in contrast to the MFT
results where $\min_t \alpha$ is between 0.633 for $L = 2400$ bp and
0.617 for $L = 23100$ bp.  Though there are large uncertainties in the
experimental data for $t < 0.01$ ms (on average 50\% for $L=2400$ bp,
going down to 10\% for $L=23100$ bp), the rough trend in the local
slope appears to show a rapid increase in $\alpha(t)$ as $t$ is
decreased.  This rapid crossover again contrasts with the MFT curve,
where the increase in the local slope is more gradual.  The heuristic
scaling results support the MFT: with the crossover exponents and
fitting constants set at the values shown below Eq.~\eqref{eq:s4}, the
heuristic $\langle r^2(t) \rangle$ almost perfectly overlaps with the
MFT curve in all three cases, and the local slopes are consequently
also very similar.  Independently, we also checked if it was possible
to find an alternative set of fitting parameters which would make the
heuristic $\langle r^2(t) \rangle$ agree with the experimental data,
but we were unable to obtain a reasonable fit.

Although the existence of an intermediate regime with sub-Zimm scaling
is found in both experiment and theory, the quantitative discrepancy
of the scaling behavior points to a gap between the experimental
system and the theoretical approaches.  Possibly, a semiflexible
polymer model based on a worm-like chain is sufficient only for
describing the large-scale motions of the DNA.  There may be some
missing elements in the theory (for example an additional degree of
freedom present in DNA, like torsional dynamics) that lead to faster
motion at shorter scales.  On the other hand, there is also the
possibility that limitations in the setup and analysis of FCS
measurements could contribute to the discrepancy.  Deviations from the
assumed Gaussian profile of the confocal detection volume and
uncertainties in the diffusion coefficient of the rhodamine molecule
used to calibrate the shape of this volume have been shown by
alternative methods like two-focus FCS to lead to substantial
systematic errors in the single-focus setup~\cite{Dertinger}.  The
uncertainties in the FCS analysis are highlighted by the differing
results produced by independent studies of similar double-stranded DNA
systems: one yielding a substantial sub-Zimm regime~\cite{Shusterman},
with local exponents near the Rouse limit, and others giving a smaller
deviation below the Zimm value over shorter time
ranges~\cite{Petrov,Winkler}, as fitted by the HWR model.  Regardless
of these issues, there is one aspect in which both the experimental
and theoretical approaches agree: an intermediate dynamical regime is
present in the end-monomer MSD results, and this regime shows sub-Zimm
scaling for long enough chains.

\section{Discussion}\label{discussion}

\begin{figure}
\begin{center}
\includegraphics*{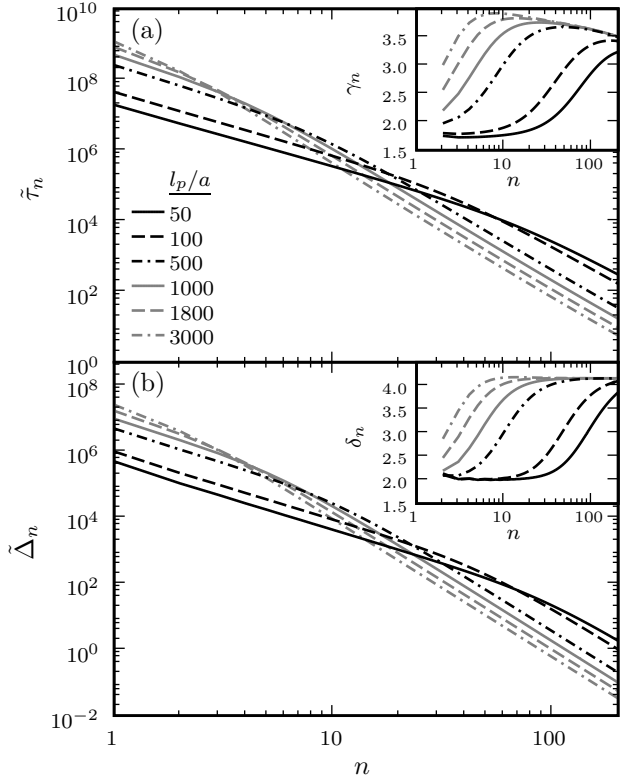}
\end{center}
\caption{The values of (a) $\tilde{\tau}_n = \tau_n k_B T \mu_0 /a^2$,
  (b) $\tilde\Delta_n = \Delta_n/a^2$ as a function of mode number $n$
  for a semiflexible polymer of length $L = 12000a$ with various $l_p
  = 50a-3000a$, calculated using MFT.  The parameters $\Delta_n$ and
  relaxation times $\tau_n$ determine the subdiffusive behavior of the
  end-monomer MSD through Eq.~\eqref{eq:t25}.  The insets show the
  effective local exponents describing the scaling of $\tau_n$ and
  $\Delta_n$ with $n$: $\gamma_n = -d\log \tau_n /d\log n$, $\delta_n =
  -d\log \Delta_n/d\log n$.}
\label{fig:taudelta}
\end{figure}

To understand the intermediate dynamical regime in more detail, and to
see where the deviations from the Zimm model arise, let us analyze the
behavior of Eq.~\eqref{eq:t25}, whose general form is shared by any
theory that expresses the end-monomer MSD in terms of contributions
from normal modes.  Both $\tau_n$ and $\Delta_n$ decrease
approximately as power laws in $n$, and we can describe this decrease
through effective local exponents $\gamma_n$, $\delta_n > 1$:
\begin{equation}\label{eq:r1}
\gamma_n = - \frac{d\log \tau_n}{d\log n}, \qquad \delta_n = -  \frac{d\log \Delta_n}{d\log n}\,.
\end{equation}
Several examples of $\tau_n$ and $\Delta_n$ are plotted in
Fig.~\ref{fig:taudelta} for $L=12000a$ and $l_p = 50a-3000a$.  The
variation of $\gamma_n$ and $\delta_n$ with $n$ is shown in the
insets.  The behavior of the MSD slope $\alpha(t)$ can be directly
related to these local exponents.  Let us assume that within some time
range $\tau_{n_2} \ll t \ll \tau_{n_1}$, $n_2 \gg n_1$, the associated
normal mode exponents are approximately constant: $\gamma_n \approx \gamma$,
$\delta_n \approx \delta$ for $n_1 < n < n_2$.  Then the dominant
contribution to the subdiffusive behavior at times $t$ within this
range is given by:
\begin{equation}\label{eq:r2}
\begin{split}
\langle r^2(t) \rangle &\approx \sum_{n=n_1}^{n_2} \Delta_{n} (1-e^{-t/\tau_{n}})\\
&\approx \Delta_{n_1}n_1^\delta \int_{n_1}^\infty dn\:n^{-\delta}(1-e^{-t n^\gamma/\tau_{n_1} n_1^\gamma})\\
&= \Delta_{n_1} n_1 \left(\frac{1}{\delta-1} + \frac{1}{\gamma}E_{1+\frac{\delta-1}{\gamma}}(t/\tau_{n_1})\right).
\end{split}
\end{equation}
In the second line we extended the upper limit of the integration from
$n_2$ to $\infty$ using the fact that $t \gg \tau_{n_2}$, $n_2 \gg
n_1$, and in the third line $E_\nu(x)$ denotes the exponential
integral function, $E_\nu(z) \equiv \int_1^\infty dt\,\exp(-zt)/t^\nu$.
For $t \ll \tau_{n_1}$, Eq.~\eqref{eq:r2} can be expanded to the
leading order as
\begin{equation}\label{eq:r3}
\langle r^2(t) \rangle \approx
\frac{\Delta_{n_1} n_1}{\delta-1}\Gamma\left(\frac{1-\delta+\gamma}{\gamma}\right) \left(\frac{t}{\tau_{n_1}}\right)^{\frac{\delta-1}{\gamma}}\,.
\end{equation}
This implies that for $\tau_{n_2} \ll t \ll \tau_{n_1}$ the local
slope is given by $\alpha(t) \approx (\delta-1)/\gamma$.  As can be
seen in the insets of Fig.~\ref{fig:taudelta}, there are two distinct
regimes for $\gamma_n$ and $\delta_n$: one for modes $n \ll L/l_p$,
corresponding to length scales greater than $l_p$, and another for
modes $n \gg L/l_p$, corresponding to length scales smaller than
$l_p$.  A continuous crossover occurs from one regime to the other
for $n \sim \text{O}(L/l_p)$.  These two regimes in turn lead to
differing behaviors for $\alpha(t)$.  We will consider each regime
separately, focusing on earlier predictions for each case and how they
compare to the present results.

Modes with $n \ll L/l_p$ correspond to internal polymer dynamics on
length scales between $l_p$ and $L$, and this is precisely the
intermediate dynamical regime that we have mentioned earlier.  In the
simplest analysis, ignoring hydrodynamical effects, these modes should
be described by the Rouse model, particularly in the flexible limit of
small $n$ where the length scales are much greater than $l_p$.  The
Rouse theory yields the following expressions for $\tau_n$ and
$\Delta_n$~\cite{DoiEdwards},
\begin{equation}\label{eq:r3b}
\tau^\text{Rouse}_n = \frac{L^2 b^2}{12\pi^2 k_B T \mu_0
  a^2} n^{-2}, \quad \Delta^\text{Rouse}_n = \frac{2L b^2}{a\pi^2}
n^{-2}\,.
\end{equation}
where $b$ is the Kuhn length, $b = \sqrt{r^2_\text{ee}(L) /M}$.  The
local exponents in the Rouse model are constants: $\delta_n = 2$ and
$\gamma_n = 2$.  Using Eq.~\eqref{eq:r3}, we find the following
asymptotic behavior for the end-monomer MSD:
\begin{equation}\label{eq:r3c}
\begin{split}
\langle r^2(t) \rangle &\approx \Delta^{\text{Rouse}}_1 \Gamma(1/2) (t/\tau_1^{\text{Rouse}})^{1/2}\\
&= \left(\frac{48 b^2 k_B T \mu_0}{\pi}\right)^{1/2} t^{1/2}\,.
\end{split}
\end{equation}
This is the origin of the Rouse scaling result $\alpha(t) = 1/2$.  In
the presence of hydrodynamic interactions, the Zimm model is expected
to hold, with Eq.~\eqref{eq:r3b} modified as:~\cite{DoiEdwards}
\begin{equation}\label{eq:r4c}
\tau^\text{Zimm}_n = \frac{b^3(L/\pi a)^{3/2}}{12\sqrt{6} k_B T \mu_0
  a} n^{-3/2}, \quad \Delta^\text{Zimm}_n = \Delta^\text{Rouse}_n\,.
\end{equation}
Here the local exponent $\gamma_n$ is 3/2, and thus the asymptotic
behavior of the MSD becomes:
\begin{equation}\label{eq:r5}
\begin{split}
\langle r^2(t)\rangle \approx \frac{12 \Gamma(1/3)(2 a k_B T \mu_0)^{2/3}}{\pi} t^{2/3}\,,
\end{split}
\end{equation}
leading to the Zimm scaling $\alpha(t) = 2/3$.

The MFT calculations, however, give a different picture, deviating
from the Zimm result.  Consider the chain in Fig.~\ref{fig:taudelta}
closest to the flexible limit: $L=12000a$, $l_p=50a$.  The exponents
$\gamma_n$ and $\delta_n$ are approximately constant for $n \lesssim
10$, but are shifted from the Zimm values: $\gamma \approx 1.74$,
$\delta \approx 2.04$ averaged over $n=1-10$, giving $\alpha =
(\delta-1)/\gamma \approx 0.60$.  Indeed in the corresponding local
slope curve plotted in Fig.~\ref{fig:mft_results}(a) the $\alpha(t)$
value is nearly constant over the time scales associated with these
modes ($t/\tau_1 \approx 0.018-1$), reaching a minimum of 0.617.  It
is these shifts in $\gamma_n$ and $\delta_n$ from the Zimm theory
predictions that lead to an intermediate dynamical regime for longer
chains where $\alpha(t) < 2/3$.

To get an analytical estimate for these shifts within the framework of
the MFT theory, one can approximately evaluate the integrals for the
interaction matrix elements $H_{nm}$ in Eq.~\eqref{eq:t22}, and
account for the effects of the off-diagonal elements using
perturbation theory.  The details of the approximation can be found in
Appendix B.  For $n \ll L/l_p$ the results are:
\begin{equation}\label{eq:r5a}
\begin{split}
\gamma_n &\approx \frac{3p_1(n) + 5.07p_2(n)\sqrt{\frac{n l_p}{L}}K\left(\frac{6a^2}{l_p^2}\right)}{2 p_3(n) + 2.89 p_4(n)\sqrt{\frac{n l_p}{L}}K\left(\frac{6a^2}{l_p^2}\right)},\quad \delta_n \approx \frac{2 p_5(n) + 0.0393 p_6(n)\sqrt{\frac{l_p}{n L}}K\left(\frac{6a^2}{l_p^2}\right)}{p_7(n) + 0.0157 p_8(n)\sqrt{\frac{l_p}{n L}}K\left(\frac{6a^2}{l_p^2}\right)},\\
\end{split}
\end{equation}
where $K(x) \equiv \sqrt{\frac{6}{\pi }} \left(E_1(x)-
E_1(3/2)\right)-4 \sqrt{\frac{3}{\pi }}+2$, and
$p_1(n),\ldots,p_{8}(n)$ are polynomials in $n$ of the form $p_i(n) =
1+a_i/n+b_i/n^2 + \cdots$, with coefficients $a_i$ and $b_i$ given in
Table 1 of Appendix B.  For $L \gg a$ and $l_p$ approaching the
flexible limit, $l_p \to 2a$, the first terms in the numerators and
denominators of the $\gamma_n$ and $\delta_n$ expressions dominate,
and thus there is a range of modes $1 \ll n \ll L/l_p$ where $\gamma_n
\approx 3/2$ and $\delta_n \approx 2$, in agreement with the expected
Zimm scaling for a flexible chain.  However, for a semiflexible chain
where $l_p \gg a$, corrections to the Zimm values become more
important.  Using the fact that $E_1(z) \approx -\log z$ as $z \to 0$,
the $K(6a^2/l_p^2)$ terms in the $\gamma_n$ expression lead to a
positive shift of order $(nl_p/L)^{1/2} \log(l_p/a)$.  For $\delta_n$
the shift upward is smaller, of order $(l_p/n L)^{1/2} \log(l_p/a)$.
These corrections due to semiflexibility are evident in the exact
numerical results for chains of length $L=12000a$ shown in the insets
of Fig.~\ref{fig:taudelta}, particularly for $l_p = 50a$ and $100a$
where a $n \ll L/l_p$ regime is identifiable.  As expected from the
analytical approximation, the deviation in $\gamma_n$ from the Zimm
value is more significant than that of $\delta_n$.  In fact the
averages of $\gamma_n$ and $\delta_n$ for $n=1-10$ from the
approximate expressions in Eq.~\eqref{eq:r5a} are 1.68 and 2.06
respectively, comparable to the numerical results 1.74 and 2.04 quoted
above.

In the other regime, for modes with $n \gg L/l_p$, the oscillations
are at length scales smaller than $l_p$, where the rigidity of the
chain is the dominating factor.  For this case it is easiest to
consider first the MFT in the absence of hydrodynamic interactions,
and then see how the final results are modified when the interactions
are included.  In the free-draining limit, the interaction matrix
$H_{nm} = 2a \mu_0 \delta_{nm}$, and for large $n$ the constants $\alpha_n$
in Eqs.~\eqref{eq:t19}-\eqref{eq:t20} are approximately $\alpha_n
\approx n \pi /L$~\cite{HWR2}.  With these simplifications we find
\begin{equation}\label{eq:r6}
\tau_n \approx \frac{L^4}{3a\mu_0
  l_p k_B T \pi^4}n^{-4},\quad \Delta_n \approx \frac{16 L^3}{l_p \pi^4}n^{-4}\,,
\end{equation}
for $n \gg L/l_p$.  Thus $\delta_n = \gamma_n=4$.  Plugging these
results into Eq.~\eqref{eq:r3} gives
\begin{equation}\label{eq:r7}
\langle r^2(t)\rangle \approx \frac{16 \Gamma(1/4)}{\pi} \frac{(a
  \mu_0 k_B T)^{3/4}}{(3 l_p)^{1/4}} t^{3/4}
\end{equation}
for the end-monomer MSD.  The scaling $\langle r^2(t) \rangle \sim
t^{3/4}$ is a well-known property of monomer motion in the stiff-rod
limit, as seen in
theory~\cite{HWR2,Kroy,Granek,FargeMaggs,Morse,GittesMackintosh,RubinsteinColby},
simulations~\cite{Everaers,Dimitrakopoulos1,Dimitrakopoulos2}, and
experiments~\cite{Bernheim,Sackmann,Weitz1,Weitz2,Wirtz,Caspi}.
Though hydrodynamic effects are typically expected to induce only weak
logarithmic corrections in this limit, we find that including these
effects in the MFT does have an observable consequence.  In the insets
of Fig.~\ref{fig:taudelta} all the $\gamma_n$ and $\delta_n$ curves
appear to overlap for $n \gg L/l_p$, but their values are shifted away
from 4: $\gamma_n$ gradually decreases with $n$, varying between 3.9
and 3.5 in the range shown, and $\delta_n \approx 4.15-4.2$.  The
behavior of $\gamma_n$ and $\delta_n$ lead to $\alpha(t) > 3/4$ in
this regime, as is seen most clearly in the large $l_p$ results in
Fig.~\ref{fig:mft_results}(a), which exhibit a broad region where the
$\alpha(t)$ curves converge over the range $0.8-0.85$ for $t \ll
\tau_p$.

The MFT
analytical estimate for the $n \gg L/l_p$ case, using the
approximation detailed in Appendix B, gives:
\begin{equation}\label{eq:r7b}
\gamma_n \approx 4 + \frac{12}{A-12 \log\left(\frac{L}{a n \pi}\right)}, \quad
\delta_n \approx 4 + \frac{24}{\left(3+A-12 \log\left(\frac{L}{a n \pi}\right)\right)\left(5+A-12 \log\left(\frac{L}{a n \pi}\right) \right)}.
\end{equation}
where the constant $A = 18 \gamma -2 \sqrt{6 \pi }+6
E_1\left(\frac{3}{2}\right)+6 \log (6) \approx 13.06$ and $\gamma
\approx 0.5772$ is Euler's constant.  The functional forms for
$\gamma_n$ and $\delta_n$ in Eq.~\eqref{eq:r7b}, independent of $l_p$,
describe the curves toward which all the $\gamma_n$ and $\delta_n$
results in the insets of Fig.~\ref{fig:taudelta} converge for
sufficiently large $n$, with $\gamma_n$ shifted below 4 and $\delta_n$
shifted above 4.  The gradual decrease in $\gamma_n$ with $n$ is
similar to an earlier theoretical approach where hydrodynamics was
explicitly considered: in Ref.~\citen{Granek} the relaxation times for
a stiff-rod were found to scale like $\tau_n \propto n^{-4}/\log(L/a
n \pi)$, corresponding to $\gamma_n = 4-1/\log(L/a n \pi)$.

In a more intuitive fashion, the heuristic scaling developed in Section II
allows to trace back the deviations from the traditional 
Zimm and worm-like-chain scaling results in Eqs.~\eqref{eq:r5} and \eqref{eq:r7}
to the slow crossovers in the diffusion and spatial size of sub-chain segments.

\section{Conclusion}\label{conclusion}

Between the flexible and stiff-rod limits hydrodynamic interactions
modify the scaling of the end-monomer MSD in ways that are not
accounted for in the Zimm model, or in earlier semiflexible polymer
theories.  In particular, there exists an intermediate dynamical
regime for sufficiently long polymers with local exponent $\alpha(t)$
between 2/3 and 1/2, the Zimm and Rouse predictions.  We have
investigated this regime through a worm-like chain model, in
conjunction with a variety of theoretical techniques: Brownian
hydrodynamics simulations for shorter chain lengths, supplemented by
mean-field theory with hydrodynamic pre-averaging for longer chains
where the simulations are not practical.  In the cases where both MFT
and numerical results are available, there is very good quantitative
agreement between them.  The two approaches are further supported by a
heuristic scaling argument that can accurately capture the trends in
$\langle r^2(t) \rangle$ and $\alpha(t)$ and that allows us to connect
the observed sub-Zimm scaling regime to previous scaling approaches
developed for the stiff-rod and the flexible-chain limits.  Note that
previous less accurate mean-field approaches that were used to analyze
the FCS data of Ref.~\citen{Petrov} give a sub-Zimm scaling range even
more pronounced than found by us.

Though the MFT and heuristics show a noticeable dip below the Zimm
exponent of 2/3 at intermediate times, they do not reach the
Rouse-like value of 1/2 seen in the experimental double-stranded DNA
results of Ref.~\citen{Shusterman}.  Comparison between the
experimental data and the theory raises a important issue: while the
long-time data, corresponding to the large-scale dynamics of the DNA,
is described surprisingly well by the MFT, serious discrepancies arise
at shorter times.  The small-scale motions revealed by experiment are
significantly faster than predicted, indicating either a deficiency in
the simple worm-like chain description or in the analysis of the FCS
measurements.  Further work is thus necessary in order to gain a
complete understanding of the monomer dynamics of DNA in solution.

\section*{Acknowledgments}

This research was supported by the Deutscher Akademischer Austausch
Dienst (DAAD) program ``Research Stays for University Academics and
Scientists'', and by the Scientific and Technical Research Council of
Turkey (T\"UB\.ITAK).  MH thanks Pamir Talazan of the Feza G\"ursey
Research Institute for assistance with the Gilgamesh computing
cluster, on which the numerical simulations were carried out.  MR
acknowledges financial support of the National Science Foundation
under grants CHE-0616925 and CBET-0609087 and the National Institutes
of Health under grant 1-R01-HL0775486A. OK acknowledges support by
Israel Science Foundation grant No.663/04.  MR, RN, and OK thank the
Kavli Institute for Theoretical Physics for initiating the
collaboration.

\section*{Appendix A:  Mean-Field Theory of an Extensible Worm-Like Chain}

As an alternative to the mean-field theory of Sec.~\ref{mft}, which
begins with the inextensible Kratky-Porod chain of Eq.~\eqref{eq:t1},
we can derive a mean-field model based on the extensible worm-like
chain Hamiltonian used in the Brownian dynamics simulations,
Eq.~\eqref{eq:n4}-\eqref{eq:n5}, thus making explicit the relationship
between the simulation and analytical results.  Ignoring the
Lennard-Jones term, the simulation Hamiltonian has the form,
\begin{equation}\label{eq:a1}
\begin{split}
U&= \frac{\gamma}{4a} \sum_{i=1}^{M-1} \left(r_{i+1,i}-2a\right)^2+\frac{\epsilon}{2a}\sum_{i=2}^{M-1}(1-\cos\theta_i)\\
&=  \frac{\Gamma }{2d} \sum_{i=1}^{M-1}(u_i - 1)^2 + \frac{\epsilon}{d} \sum_{i=2}^{M-1} \left(1-\frac{\mb{u}_i\cdot \mb{u}_{i-1}}{u_i u_{i-1}}\right)\,,
\end{split}
\end{equation}
where $d=2a$, $\Gamma = \gamma d^2$, and $\mb{u}_i =
(\mb{r}_{i+1}-\mb{r}_i)/d$.  For large $\Gamma$ (the case in the
simulations), the values of $u_i = |\mb{u}_i| \approx 1$, and we can
expand $(u_i - 1)^2 = (\sqrt{1+(u_i^2-1)}-1)^2 \approx (u_i^2-1)^2/4 +
\text{O}((u_i^2-1)^3)$.  Keeping the leading term, we rewrite
Eq.~\eqref{eq:a1} as
\begin{equation}\label{eq:a2}
U \approx \frac{\Gamma}{8d} \sum_{i=1}^{M-1} (u_i^2 -1)^2 - \frac{\epsilon}{d} \sum_{i=2}^{M-1} \mb{u}_i \cdot \mb{u}_{i-1} + \frac{\epsilon}{d}(M-2)\,.
\end{equation}
Neglecting the last term of Eq.~\eqref{eq:a2}, since it is a constant,
the chain partition function $Z$ is given by
\begin{equation}\label{eq:a3}
\begin{split}
Z &= \int \prod_{i=1}^{M-1} d\mb{u}_i \, e^{-\beta U}\\
&= \int \prod_{i=1}^{M-1} d\mb{u}_i \, e^{-\frac{\beta \Gamma}{8 d} \sum_{i=1}^{M-1} (u_i^2 -1)^2 + \frac{\beta \epsilon}{d} \sum_{i=2}^{M-1} \mb{u}_i \cdot \mb{u}_{i-1}}.
\end{split}
\end{equation}
We can rewrite the integrand of $Z$ using the relations
\begin{equation}\label{eq:a4}
e^{-\frac{\beta \Gamma}{8 d} (u_i^2 -1)^2} \propto \int_{-i\infty}^{i \infty} d\lambda_i \,e^{-\beta \lambda_i d (u_i^2 -1)+ \frac{2 \beta d^3}{\Gamma}\lambda_i^2}\,,
\end{equation}
and $\mb{u}_i \cdot \mb{u}_{i-1} = \frac{1}{2}\left(u_i^2 + u_{i-1}^2- (\mb{u}_i -\mb{u}_{i-1})^2\right)$, where we have introduced an auxiliary variable $\lambda_i$ for each $i$.  The result, up to a constant prefactor,  is
\begin{equation}\label{eq:a5}
Z = \int_{-i\infty}^{i\infty} \prod_{i=1}^{M-1} d\lambda_i \, e^{-\beta F(\{\lambda_i\})}\,,
\end{equation}
with 
\begin{equation}\label{eq:a6}
F(\{\lambda_i\}) = -\beta^{-1} \log \int \prod_{i=1}^{M-1} d\mb{u}_i \, e^{-\beta U(\{\lambda_i\})}\,,\\
\end{equation}
and
\begin{equation}\label{eq:a6b}
\begin{split}
U(\{\lambda_i\}) =& d \sum_{i=2}^{M-2} \left(\lambda_i - \frac{\epsilon}{d^2}\right) u_i^2
+ \frac{\epsilon}{2d} \sum_{i=2}^{M-1} (\mb{u}_i - \mb{u}_{i-1})^2\\
&+d\left(\lambda_1- \frac{\epsilon}{2d^2} \right)u_1^2+d\left(\lambda_{M-1}- \frac{\epsilon}{2d^2} \right)u_{M-1}^2\\
&-d \sum_{i=1}^{M-1} \lambda_i - \frac{2d^3}{\Gamma}\sum_{i=1}^{M-1}\lambda_i^2 \,.
\end{split}
\end{equation}
To derive a mean-field model we can now apply a stationary phase
approximation analogous to the one used in Sec.~\ref{mft}
\cite{HaThirumalai1,HaThirumalai2}:
\begin{equation}\label{eq:a7}
Z = \int_{-i\infty}^{i\infty} \prod_{i=1}^{M-1} d\lambda_i \, e^{-\beta F(\{\lambda_i\})} \approx e^{-\beta F(\{\lambda^\text{cl}_i\})}\,,
\end{equation}
where $\{\lambda^\text{cl}_i\}$ satisfy
\begin{equation}\label{eq:a8}
\left.\frac{\partial F}{\partial \lambda_i}\right|_{\{\lambda_i = \lambda^\text{cl}_i\}} = 0, \qquad i=1,\ldots,M-1\,.
\end{equation}
From symmetry, we know $\lambda^\text{cl}_{i}$ must have the property
$\lambda^\text{cl}_{i}=\lambda^\text{cl}_{M-i}$ for all $i$, and thus
we can write the solution to Eq.~\eqref{eq:a8} in the form
\begin{equation}\label{eq:a9}
\begin{split}
\lambda_i^\text{cl} &= \nu_i + \frac{\epsilon}{d^2}, \qquad i=2,\ldots,M-2\,,\\
\lambda_1^\text{cl} &= \lambda_{M-1}^\text{cl} = \frac{\nu_0}{d} + \frac{\epsilon}{2d^2}\,,
\end{split} 
\end{equation}
for some set of values $\{\nu_i\}$ and $\nu_0$, where $\nu_i = \nu_{M-i}$.  This
yields a mean-field free energy
\begin{equation}\label{eq:a10}
F_\text{MF} \equiv   F(\{\lambda^\text{cl}_i\}) = -\beta^{-1} \log \int \prod_{i=1}^{M-1} d\mb{u}_i \, e^{-\beta U_\text{MF}}\,,
\end{equation}
with
\begin{equation}\label{eq:a10b}
\begin{split}
U_\text{MF} = &  d\sum_{i=2}^{M-2} \nu_i u_i^2
+ \frac{\epsilon}{2d} \sum_{i=2}^{M-1} (\mb{u}_i - \mb{u}_{i-1})^2\\
&+\nu_0(u_1^2+u_{M-1}^2) -d \sum_{i=2}^{M-2} \nu_i -2\nu_0\\
& - \frac{2d^3}{\Gamma}\sum_{i=2}^{M-2}\left(\nu_i+\frac{\epsilon}{d^2}\right)^2 -\frac{4 d^3}{\Gamma}\left(\frac{\nu_0}{d}+\frac{\epsilon}{2d^2}\right)^2\,.
\end{split}
\end{equation}
In the continuum limit $d \to 0$, $M \to \infty$, $M d \to L$ and we
replace $\mb{u}_i$, $\nu_i$ by continuous functions $\mb{u}(s)$,
$\nu(s)$ of the contour variable $s$.  Assuming the Hamiltonian
parameters $\epsilon$ and $\Gamma$ remain fixed in this limit, we have
\begin{equation}\label{eq:a11}
\begin{split}
F_\text{MF}  =& -\beta^{-1} \log \int {\cal D}\mb{u} \, e^{-\beta U_\text{MF}}\,,\\
U_\text{MF} =& \int_{-L/2}^{L/2} ds\, \nu(s) \mb{u}^2(s) + \frac{\epsilon}{2} \int_{-L/2}^{L/2} ds\, \left(\frac{\partial\mb{u}(s)}{\partial s} \right)^2\\
&+\nu_0(\mb{u}^2(L/2)+\mb{u}^2(-L/2))-\int_{-L/2}^{L/2} ds\,\nu(s)\\
&-2\nu_0-\frac{4\epsilon}{\Gamma}\int_{-L/2}^{L/2} ds\,\nu(s)-\frac{4\nu_0 \epsilon}{\Gamma} -C\,,
\end{split}
\end{equation}
where $C = \lim_{d \to 0} (2L\epsilon^2/\Gamma d^2 + \epsilon^2/\Gamma
d)$ is an infinite constant independent of $\nu(s)$ and $\nu_0$.  The stationary point condition Eq.~\eqref{eq:a8} becomes
\begin{equation}\label{eq:a11b}
\frac{\delta F_\text{MF}}{\delta \nu(s)} = 0, \qquad \frac{\partial F_\text{MF}}{\partial \nu_0} = 0\,.
\end{equation}
Physically Eq.~\eqref{eq:a11b} implies the following constraints:
\begin{equation}\label{eq:a11c}
\begin{split}
&\langle \mb{u}^2(s) \rangle = 1+ \frac{4\epsilon}{\Gamma}\,, \qquad -L/2 < s < L/2\,,\\
&\langle \mb{u}^2(-L/2) \rangle = \langle \mb{u}^2(L/2) \rangle = 1+ \frac{2\epsilon}{\Gamma}\,,
\end{split}
\end{equation}
where $\langle \: \cdot \: \rangle$ denotes the thermal
average with respect to the Hamiltonian $U_\text{MF}$.  As expected,
the magnitude of the tangent vector fluctuations become smaller as the
extensibility parameter $\Gamma$ increases, going to the limit
$\langle \mb{u}^2(s) \rangle = 1$ for all $s$ when $\Gamma \to
\infty$.  As will be seen below, in this limit the present theory
reproduces the results of Sec.~\ref{mft}.

\begin{figure}
\centering \includegraphics*[scale=1]{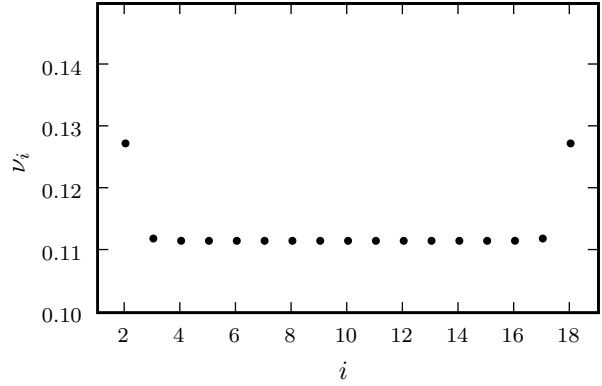}
\caption{Numerical solution to the stationary point condition
  Eq.~\eqref{eq:a8} for a system with $M=20$, $\epsilon = 10 \: k_BTd$,
  and $\Gamma = 400 \: k_BTd$.  The solution is expressed in terms of
  $\nu_i = \lambda_i^\text{cl} - \epsilon/d^2$ for $i=2,\ldots,M-2$,
  in units of $k_B T /d$.  The value of $\nu_0/d = \lambda_1^\text{cl}
  - \epsilon/2d^2 = \lambda_{M-1}^\text{cl} - \epsilon/2d^2$ is $0.7831
  \: k_B T /d$.}\label{fig:app1}
\end{figure}

Unfortunately Eq.~\eqref{eq:a11b} is not analytically tractable,
because the $F_\text{MF}$ given by Eq.~\eqref{eq:a11} cannot be
evaluated in closed form for an arbitrary function $\nu(s)$.  On the
other hand, the stationary point condition in the discrete system,
Eq.~\eqref{eq:a8} for $F$ in Eq.~\eqref{eq:a6}, can be solved
numerically for small $M$, and a representative set of $\nu_i$ are
shown in Fig.~\ref{fig:app1} (for $M=20$, $\epsilon = 10 k_BTd$,
$\Gamma = 400 k_BTd$).  For large $\Gamma \gg \epsilon$, the $\nu_i$ are
nearly constant for $2 \le i \le M-2$, and we can use this fact
to make the following approximation in the continuum limit: replace
$\nu(s)$ by a constant $\nu$ in the Hamiltonian $U_\text{MF}$ of
Eq.~\eqref{eq:a11}.  Thus the first part of Eq.~\eqref{eq:a11b} becomes
$\partial F_\text{MF}/\partial \nu = 0$, implying a global
constraint
\begin{equation}\label{eq:a11d}
\int_{-L/2}^{L/2} ds\,\langle \mb{u}^2(s) \rangle = \left(1+\frac{4\epsilon}{\Gamma}\right)L\,,
\end{equation}
instead of the local constraint $\langle \mb{u}^2(s) \rangle = 1+4\epsilon/\Gamma$ in Eq.~\eqref{eq:a11c}.

This approximation, which becomes exact when
$\Gamma \to \infty$, allows us to find a closed form expression for
$F_\text{MF}$.  Using a mapping of the first two terms of
$U_\text{MF}$ to the quantum mechanical harmonic oscillator (with mass
$m = \epsilon$ and frequency $\omega = \sqrt{2\nu/\epsilon}$)
\cite{HaThirumalai1}, the path integral for $F_\text{MF}$ can be
evaluated, giving the free energy
\begin{equation}\label{eq:a12}
\begin{split}
&F_\text{MF} = -L\nu -2\nu_0 -\frac{4L \nu \epsilon}{\Gamma}-\frac{4\nu_0 \epsilon}{\Gamma}\\
&\quad -\frac{3}{2\beta}\left(\log\left[\beta\sqrt{\nu \epsilon}\,\text{csch}\left(L \sqrt{\frac{2\nu}{\epsilon}} \right) \right]\right.\\
&\quad -\left.\log\left[\frac{\beta^2(2\nu_0^2 + \nu \epsilon)}{2}+\beta^2 \nu_0 \sqrt{2\nu \epsilon}\, \text{coth}\left(L \sqrt{\frac{2\nu}{\epsilon}} \right)\right]\right)\,,
\end{split}
\end{equation}
up to an additive constant.  Using Eq.~\eqref{eq:a12}, the stationary
point condition $\partial F_\text{MF}/\partial \nu = \partial
F_\text{MF}/\partial \nu_0 = 0$ can be solved numerically for $\nu$
and $\nu_0$ given $L$, $\epsilon$, and $\Gamma$.  For $L \gg \epsilon$, the
condition takes the simple limiting form,
\begin{equation}\label{eq:a14}
\begin{split}
\sqrt{\frac{\nu \epsilon}{2}} &= \frac{3}{4}k_B T \frac{\Gamma}{\Gamma+4\epsilon}\,,\\
\nu_0 &= \frac{3}{4}k_B T \frac{\Gamma(\Gamma+6\epsilon)}{(\Gamma+2\epsilon)(\Gamma+4\epsilon)}\,.
\end{split}
\end{equation}
When $\Gamma \to \infty$, Eq.~\eqref{eq:a14} reduces to
Eq.~\eqref{eq:t7} in Sec.~\ref{mft}, and this is generally true of the
stationary point condition for any $L$.

The equilibrium properties of the chain described by the Hamiltonian
$U_\text{MF}$ can be calculated with a similar approach to the one
used in Ref.~\citen{HWR1}, where distribution functions were derived
for the $\Gamma = \infty$ mean-field model.  The main result is
$G(s,s^\prime;\mb{x};\mb{u},\mb{u}^\prime)$, the probability density
for finding two points on the chain at $s$ and $s^\prime$ with
$s^\prime > s$, having spatial separation $\mb{r}(s^\prime) -
\mb{r}(s) = \mb{x}$, and tangent vectors $\mb{u}(s) = \mb{u}$,
$\mb{u}(s^\prime) = \mb{u}^\prime$.  The full expression for this
probability is
\begin{equation}\label{eq:a15}
\begin{split}
&G(s,s^\prime;\mb{x};\mb{u},\mb{u}^\prime)=\\
&\quad \left(\frac{A(s,s^\prime)}{4\pi^3 B_4(s^\prime - s)} \right)^{3/2} \exp\Biggl[-B_1(s^\prime-s)(\mb{u}^2+{\mb{u}^\prime}^2)\\
&\quad+ B_2(s^\prime-s) \mb{u}\cdot \mb{u}^\prime - \frac{\left(\mb{x} - B_3(s^\prime-s)(\mb{u}+\mb{u}^\prime)\right)^2}{B_4(s^\prime-s)}\\
&\quad- C(s+L/2) \mb{u}^2 - C(L/2-s^\prime) {\mb{u}^\prime}^2\Biggr]\,,
\end{split}
\end{equation}
where the functions $A(s)$, $B_i(s)$, $i=1,\ldots,4$, and $C(s)$ are
given by
\begin{equation}\label{eq:a17}
\begin{split}
A(s,s^\prime) &= 4 (B_1(s^\prime-s)+C(s+L/2))\\
&\quad\cdot(B_1(s^\prime-s)+C(L/2-s^\prime))-B_2^2(s^\prime-s)\,,\\
B_1(s) &= \frac{\beta\epsilon \omega}{2} \coth (s \omega)\,,\quad B_2(s) = \beta \epsilon \omega\, \text{csch}(s\omega)\,,\\
B_3(s) &= \frac{\epsilon \omega}{2\nu} \tanh\left(\frac{s\omega}{2} \right)\,,\\
B_4(s) &= \frac{s}{\beta\nu}-\frac{\epsilon \omega}{\beta\nu^2}\tanh\left(\frac{s\omega}{2}\right)\,,\\
C(s) &= \frac{\beta \epsilon \omega (\epsilon \omega+2\nu_0\,\text{coth}(s\omega))}{4\nu_0 + 2\epsilon \omega\,\text{coth}(s\omega)}\,,
\end{split}
\end{equation}
with $\omega = \sqrt{2\nu/\epsilon}$.

From $G(s,s^\prime;\mb{x};\mb{u},\mb{u}^\prime)$ we can calculate
other properties of the chain, for example the tangent vector
correlation function $\langle \mb{u}(s) \cdot \mb{u}(s^\prime)
\rangle$ for $s^\prime > s$,
\begin{equation}\label{eq:a18}
\begin{split}
\langle \mb{u}(s)\cdot \mb{u}(s^\prime)\rangle &= \int d^3\mb{x}\, \mb{u}(s) \cdot \mb{u}(s^\prime)\, G(s,s^\prime;\mb{x};\mb{u}(s),\mb{u}(s^\prime))\\ 
&= \frac{B_2(s^\prime - s)}{A(s^\prime,s)}\,.
\end{split}
\end{equation}
For $L \gg \epsilon$, $-L/2 \ll s,s^\prime \ll L/2$,
Eq.~\eqref{eq:a18} can be simplified using Eq.~\eqref{eq:a14} for
$\nu$ and $\nu_0$, giving
\begin{equation}\label{eq:a19}
\langle \mb{u}(s)\cdot \mb{u}(s^\prime)\rangle = \left(1+\frac{4\epsilon}{\Gamma}\right)\exp\left(-\frac{3 (s^\prime -s)\Gamma k_B T}{2\epsilon(\Gamma+4\epsilon)}\right)\,.
\end{equation}
When $\Gamma \to \infty$, the tangent correlation function reduces to
the Kratky-Porod form of Eq.~\eqref{eq:t8}, $\langle \mb{u}(s)\cdot
\mb{u}(s^\prime)\rangle = \exp(-3 (s^\prime - s) k_B T/2\epsilon) =
\exp(-(s^\prime-s)/l_p)$.

Thus we have shown that a mean-field theory based on the extensible
worm-like chain Hamiltonian used in the simulations gives results very
similar to the MFT described in Sec.~\ref{mft}, with the finite
extensibility leading to small corrections to the parameters of
$U_\text{MF}$ on the order of $\epsilon/\Gamma$.

\section*{Appendix B:  Analytical Approximation for $\gamma_n$ and $\delta_n$}

In order to derive analytical expressions for the exponents $\gamma_n$
and $\delta_n$ from the MFT theory, we make several approximations in
the derivation described in Sec.~\ref{mft}.  Since the off-diagonal
elements of the interaction matrix $H_{nm}$ defined by
Eq.~\eqref{eq:t22} are smaller than the diagonal ones, we will treat
them as a perturbation.  To first order in the perturbation
expansion, we can write the following expressions for $\tau_n$,
$\Theta_n$, and $\Psi_n(L/2)$ when $n>0$:
\begin{equation}\label{eq:ba1}
\begin{split}
\tau_n &= \Lambda_n^{-1} \approx \lambda_n^{-1}H_{nn}^{-1}- H_{nn}^{-2}\lambda_n^{-1}\sum_{m\ne n} \frac{H_{nm}^2\lambda_m}{H_{nn}\lambda_n - H_{mm} \lambda_m},\\
\Theta_n &\approx H_{nn} + 2\sum_{m\ne n} \frac{H_{n
m}^2 \lambda_m}{H_{nn}\lambda_n - H_{mm}\lambda_m},\\
\Psi_n(L/2) &= \sum_m \psi_m(L/2)(C^{-1})_{mn} \approx \psi_n(L/2) + \sqrt{\frac{1}{L}} \frac{H_{0n}}{H_{nn}}+\sum_{m\ne n} \psi_m(L/2) \frac{H_{nm} \lambda_n }{H_{nn}\lambda_n - H_{mm}\lambda_m}.
\end{split}
\end{equation}
From these expressions one can also calculate $\Delta_n = 6k_B T
\tau_n \Theta_n \Psi_n^2(L/2)$.  The double integral for $H_{nm}$ in
Eq.~\eqref{eq:t22} can be rewritten in terms of new variables
$h=s-s^\prime$ and $w=s+s^\prime$ as follows:
\begin{equation}\label{eq:ba2}
\begin{split}
H_{nm} &= 2a\mu_0\delta_{nm} + \sqrt{\frac{6}{\pi}}a\mu_0 \int_{-L/2}^{L/2}ds \int_{-L/2}^{L/2}ds^\prime\,\psi_n(s) \frac{\Theta(|s-s^\prime|-2a)}{\sqrt{\sigma(|s-s^\prime|)}} \exp\left(-\frac{6a^2}{\sigma(|s-s^\prime|)}\right) \psi_m(s^\prime)\,\\
&= 2a\mu_0\delta_{nm} + \sqrt{\frac{6}{\pi}}a\mu_0 \int_{2a}^{L} dh \int_{-L+h}^{L-h} dw\,\psi_n\left(\frac{h+w}{2}\right) \frac{1}{\sqrt{\sigma(h)}} \exp\left(-\frac{6a^2}{\sigma(h)}\right) \psi_m\left(\frac{w-h}{2}\right)\,.
\end{split}
\end{equation}
Since $\sigma(h) = 2l_p h - 2l_p^2 (1-e^{-h/l_p})$ can be approximated
as $\sigma(h) \approx h^2$ for $h \ll l_p$ and $\sigma(h) \approx 2l_p
h$ for $h \gg l_p$, we can split up the $h$ integral above into two
pieces:
\begin{equation}\label{eq:ba3}
H_{nm} \approx  2a\mu_0\delta_{nm} + a \mu_0 (I^{(1)}_{nm}+I^{(2)}_{nm}),
\end{equation}
where
\begin{equation}\label{eq:ba4}
\begin{split}
I^{(1)}_{nm} &= \sqrt{\frac{6}{\pi}} \int_{2a}^{l_p} dh \,\frac{1}{h} \exp\left(-\frac{6a^2}{h^2}\right) \int_{-L+h}^{L-h} dw\,\psi_n\left(\frac{h+w}{2}\right)\psi_m\left(\frac{w-h}{2}\right),\\
I^{(2)}_{nm} &= \sqrt{\frac{6}{\pi}} \int_{l_p}^{L} dh \,\frac{1}{\sqrt{2l_p h}} \exp\left(-\frac{3a^2}{l_p h}\right) \int_{-L+h}^{L-h} dw\,\psi_n\left(\frac{h+w}{2}\right)\psi_m\left(\frac{w-h}{2}\right).
\end{split}
\end{equation}
To complete the approximation, we will estimate these integrals in the
two mode regimes discussed in Sec.~\ref{discussion}, one for the case
$n \ll L/l_p$, the other for $n \gg L/l_p$.  Since the biggest
perturbation contributions in Eq.~\eqref{eq:ba1} for $\tau_n$,
$\Theta_n$, and $\Psi_n(L/2)$ come from states with $m$ in the
vicinity of $n$, it is sufficient to consider matrix elements $H_{mn}$
for $m$ in the same mode regime as $n$.

\subsubsection*{B.1. $n,m \ll L/l_p$ regime}

In the limit of long chain lengths, where $L \gg l_p,\: a$, the
functions $\psi_n(s)$ and constants $\alpha_n$, $\lambda_n$ for $0<n \ll L/l_p$
in Eqs.~\eqref{eq:t19} and \eqref{eq:t20} simplify to:
\begin{equation}\label{eq:b1}
\begin{split}
\psi_n(s) &\approx \begin{cases} (-1)^{(n-1)/2}\sqrt{\frac{2}{L}} \sin \left(\frac{\pi n
s}{L}\right) & n\: \text{odd},\\
(-1)^{n/2}\sqrt{\frac{2}{L}} \cos \left(\frac{\pi n s}{L}\right) & n\:
\text{even},
\end{cases}\\ 
\alpha_n &\approx \frac{\pi n}{L}, \quad \lambda_n \approx \frac{3k_B T \pi^2 n^2}{2l_p L^2}\,.
\end{split}
\end{equation}
Due to the symmetry of the $\psi_n(s)$ functions, the matrix elements
$H_{nm}$ are non-zero only when $n$ and $m$ are both odd or both even,
so the perturbation expansions in Eq.~\eqref{eq:ba1} can be done
independently for even and odd states.  For simplicity, we will assume
$n$ and $m$ are odd for the rest of the derivation.  Carrying out the
analogous approximation for even $n$, $m$, will lead to qualitatively
similar final expressions for $\delta_n$ and $\gamma_n$, with slight
shifts in the numerical coefficients.

Plugging Eq.~\eqref{eq:b1} into Eq.~\eqref{eq:ba4}, we can approximately
evaluate the integrals in the large $L$ limit:
\begin{equation}\label{eq:ba5}
\begin{split}
I^{(1)}_{nm} &\approx \begin{cases} \frac{4\sqrt{6}}{L\sqrt{\pi}}\left[2ae^{-3/2}-l_p e^{-6a^2/l_p^2}+a\sqrt{6\pi}\,\text{erf}(\sqrt{3/2})-a\sqrt{6\pi}\,\text{erf}(\sqrt{6}a/l_p)\right] & n\ne m,\\
\sqrt{\frac{6}{\pi}}\left[E_1\left(6a^2/l_p^2\right)-E_1\left(3/2\right) \right] & n=m,\end{cases}\\
I^{(2)}_{nm} &\approx \begin{cases} -\sqrt{\frac{L}{l_p}}\frac{2\sqrt{6}}{(n+m)(\sqrt{n}+\sqrt{m})\pi^{3/2}} & n\ne m,\\  \sqrt{\frac{6L}{\pi l_p}}\left(\frac{1}{n^{1/2}} - \frac{1}{2\pi n^{3/2}} \right)- 2\sqrt{\frac{12}{\pi}} & n=m. \end{cases}
\end{split}
\end{equation}
Plugging these results into Eq.~\eqref{eq:ba3} for $H_{nm}$, we can
also estimate the sums involved in the perturbation expansion of
Eq.~\eqref{eq:ba1}:
\begin{equation}\label{eq:ba6}
\begin{split}
\sum_{m\ne n} \frac{H_{nm}^2\lambda_m}{H_{nn}\lambda_n - H_{mm}
  \lambda_m} &\approx \sqrt{\frac{L}{l_p}}\left(\frac{-108 \pi +27 \pi ^2}{18 \sqrt{6} \pi ^{7/2} n^{3/2}}+\frac{-126+72 \pi -16 \sqrt{3} \pi }{18
   \sqrt{6} \pi ^{7/2} n^{5/2}} \right),\\
\sum_{m\ne n} \psi_m(L/2) \frac{H_{nm}\lambda_n}{H_{nn}\lambda_n - H_{mm}
  \lambda_m} &\approx \sqrt{\frac{2}{L}}\left[\frac{1}{18} \left(-9+4 \sqrt{3}\right)+\frac{-9+\left(9+\sqrt{3}\right) \pi }{27 \pi ^2 n}-\frac{72-45 \pi +4 \sqrt{3} \pi }{216 \pi ^3 n^2}\right].
\end{split}
\end{equation}
Combining the results of Eqs.~\eqref{eq:ba1}, \eqref{eq:ba3}, and
\eqref{eq:b1}-\eqref{eq:ba6}, we can derive the following expressions
for $\gamma_n$ and $\delta_n$:
\begin{equation}\label{eq:ba7}
\begin{split}
\gamma_n &= -\frac{d \log \tau_n}{d\log n} = -\frac{n}{\tau_n} \frac{d \tau_n}{dn} \approx \frac{3p_1(n) + 5.07p_2(n)\sqrt{\frac{n l_p}{L}}K\left(\frac{6a^2}{l_p^2}\right)}{2 p_3(n) + 2.89 p_4(n)\sqrt{\frac{n l_p}{L}}K\left(\frac{6a^2}{l_p^2}\right)},\\
\delta_n &= -\frac{d \log \Delta_n}{d\log n} = -\frac{n}{\Delta_n} \frac{d \Delta_n}{dn} \approx \frac{2 p_5(n) + 0.0393 p_6(n)\sqrt{\frac{l_p}{n L}}K\left(\frac{6a^2}{l_p^2}\right)}{p_7(n) + 0.0157 p_8(n)\sqrt{\frac{l_p}{n L}}K\left(\frac{6a^2}{l_p^2}\right)},\\
\end{split}
\end{equation}
where
\begin{equation}\label{eq:ba8}
K(x) \equiv \sqrt{\frac{6}{\pi }} \left(E_1(x)- E_1(3/2)\right)-4 \sqrt{\frac{3}{\pi
   }}+2,
\end{equation}
and $p_1(n),\ldots,p_{8}(n)$ are polynomials in $n$ of the form
$p_i(n) = 1+a_i/n+b_i/n^2 + \cdots$.  The first two coefficients $a_i$ and
$b_i$ are given in the following table:

\begin{table}[h]\label{tab1}
\centering {\footnotesize \begin{tabular}{c|cccccccc}
$i$ & 1 & 2 & 3 & 4 & 5 & 6 & 7 & 8\\
\hline
$a_i$ & -0.176 & -0.0919 & -0.297 & -0.148 & 0.282 & 0.608 & 0.188 & 0.435 \\
$b_i$ &  -0.00186 & -0.00504 & 0.0179 & -0.00196 & 0.0440 & 0.185 & 0.0220 & 0.103
\end{tabular}}
\caption{Coefficients $a_i$ and $b_i$ of polynomials $p_i(n) = 1+a_i/n+ b_i/n^2 + \cdots$. }
\end{table}

\subsubsection*{B.2. $n,m \gg L/l_p$ regime}

With the assumptions that $L \gg l_p$ and $l_p \gg a$, the functions
$\psi_n(s)$ and constants $\alpha_n$, $\lambda_n$ for $n \gg L/l_p$ in
Eqs.~\eqref{eq:t19} and \eqref{eq:t20} become:
\begin{equation}\label{eq:ba9}
\begin{split}
\psi_n(s) &\approx \begin{cases} (-1)^{(n-1)/2}\sqrt{\frac{2}{L}} \sin \left(\frac{\pi (2n-1)
s}{2L}\right) + \sqrt{\frac{1}{L}} \frac{\sinh\left(\pi(2n-1)s/2L\right)}{\cosh\left(\pi(2n-1)/4\right)} & n\: \text{odd},\\
(-1)^{n/2}\sqrt{\frac{2}{L}} \cos \left(\frac{\pi (2n-1) s}{2L}\right) + \sqrt{\frac{1}{L}} \frac{\cosh\left(\pi(2n-1)s/2L\right)}{\sinh\left(\pi(2n-1)/4\right)} & n\:
\text{even},
\end{cases}\\ 
\alpha_n &\approx \frac{\pi (2n-1)}{2L}, \quad \lambda_n \approx \frac{3k_B T l_p \pi^4 (2n-1)^4}{32L^4}.
\end{split}
\end{equation}
Again we will focus for simplicity on the case of odd $n$ and $m$.
For $n,m \gg L/l_p$, the first integral in Eq.~\eqref{eq:ba4}
dominates, $I_{nm}^{(1)} \gg I_{nm}^{(2)}$, so we can write $H_{nm}
\approx 2a\mu_0 \delta_{nm} + a\mu_0 I^{(1)}_{nm}$.  The integral
$I_{nm}^{(1)}$ can be approximated as:
\begin{equation}\label{eq:ba10}
I^{(1)}_{nm} \approx \begin{cases} \frac{2 \sqrt{6}}{\pi^{3/2}} \frac{\left(m^2+n^2\right) \left(\log \left(\frac{16 a^2 m n \pi ^2}{L^2}\right)+2 \gamma
   \right)-(m+n-1) (m+n)^2 \pi}{(m+n-1) (m+n) \left(m^2+n^2\right)} & n\ne m,\\
-\frac{\sqrt{6}}{n\pi^{3/2}}\left[3+ n\pi\left\{ 3\gamma+E_1\left(\frac{3}{2}\right) + 2\log\left(\frac{\sqrt{6}a\pi n}{L}\right)\right\} \right] & n=m,\end{cases}
\end{equation}
where $\gamma \approx 0.5772$ is Euler's constant.  The resulting sums
in the perturbation expansion of Eq.~\eqref{eq:ba1} are:
\begin{equation}\label{eq:ba11}
\begin{split}
\sum_{m\ne n} \frac{H_{nm}^2\lambda_m}{H_{nn}\lambda_n - H_{mm}
  \lambda_m} &\approx \frac{12\sqrt{6}}{n \sqrt{\pi } \left(3+A-12\log \left(\frac{L}{a n \pi }\right)\right)},\\ \sum_{m\ne
  n} \psi_m(L/2) \frac{H_{nm}\lambda_n}{H_{nn}\lambda_n - H_{mm}
  \lambda_m} &\approx \frac{2}{\sqrt{L}} \left(\frac{1}{3+A-12 \log\left(\frac{L}{a n \pi}\right)}\right),
\end{split}
\end{equation}
where the constant $A = 18 \gamma -2 \sqrt{6 \pi }+6
E_1\left(\frac{3}{2}\right)+6 \log (6) \approx 13.06$.  This yields
the following expressions for $\gamma_n$ and $\delta_n$:
\begin{equation}\label{eq:ba12}
\gamma_n \approx 4 + \frac{12}{A-12 \log\left(\frac{L}{a n \pi}\right)}, \quad
\delta_n \approx 4 + \frac{24}{\left(3+A-12 \log\left(\frac{L}{a n \pi}\right)\right)\left(5+A-12 \log\left(\frac{L}{a n \pi}\right) \right)}.
\end{equation}

\end{document}